\documentclass[pdflatex,sn-mathphys-num]{sn-jnl}

\usepackage{graphicx}%
\usepackage{multirow}%
\usepackage{amsmath,amssymb,amsfonts}%
\usepackage{amsthm}%
\usepackage{mathrsfs}%
\usepackage[title]{appendix}%
\usepackage{xcolor}%
\usepackage{textcomp}%
\usepackage{manyfoot}%
\usepackage{booktabs}%
\usepackage{algorithm}%
\usepackage{algorithmicx}%
\usepackage{algpseudocode}%
\usepackage{listings}%
\usepackage{subfig}

\usepackage{cleveref}
\crefname{figure}{fig.}{figs.}
\Crefname{figure}{Fig.}{Figs.}
\crefname{table}{tbl.}{tbls.}
\Crefname{table}{Tbl.}{Tbls.}
\crefname{equation}{}{}
\Crefname{equation}{Eq.}{Eqns.}
\crefname{listing}{lst.}{lsts.}
\Crefname{listing}{Lst.}{Lsts.}
\crefname{section}{\S}{\S}
\Crefname{section}{Sec.}{Secs.}
\crefname{codelisting}{\cref@listing@name}{\cref@listing@name@plural}
\Crefname{codelisting}{\Cref@listing@name}{\Cref@listing@name@plural}

\theoremstyle{thmstyleone}%
\theoremstyle{thmstyletwo}%

\theoremstyle{thmstylethree}%

\begin{document}

\title[Overcoming error-in-variable problem in data-driven model
discovery by orthogonal distance regression]{Overcoming error-in-variable problem in data-driven model
discovery by orthogonal distance regression}


\author*[1]{\fnm{Lloyd} \sur{Fung}}\email{lloyd.fung@imperial.ac.uk}

\affil*[1]{\orgdiv{Department of Aeronautics}, \orgname{Imperial College London}, \orgaddress{\street{Exhibition Road}, \city{London}, \postcode{SW7 2AZ}, \country{UK}}}


\abstract{Despite the recent proliferation of machine learning methods like SINDy that promise automatic discovery of governing equations from time-series data, there remain significant challenges to discovering models from noisy datasets. One reason is that the linear regression underlying these methods assumes that all noise resides in the training target (the regressand), which is the time derivative, whereas the measurement noise is in the states (the regressors). Recent methods like modified-SINDy and DySMHO address this error-in-variable problem by leveraging information from the model's temporal evolution, but they are also imposing the equation as a hard constraint, which effectively assumes no error in the regressand. Without relaxation, this hard constraint prevents assimilation of data longer than Lyapunov time. Instead, the fulfilment of the model equation should be treated as a soft constraint to account for the small yet critical error introduced by numerical truncation. The uncertainties in both the regressor and the regressand invite the use of orthogonal distance regression (ODR). By incorporating ODR with the Bayesian framework for model selection, we introduce a novel method for model discovery, termed ODR-BINDy, and assess its performance against current SINDy variants using the Lorenz63, R{\"o}ssler, and Van Der Pol systems as case studies. Our findings indicate that ODR-BINDy consistently outperforms all existing methods in recovering the correct model from sparse and noisy datasets. For instance, our ODR-BINDy method reliably recovers the Lorenz63 equation from data with noise contamination levels of up to \(30\%\).}




\maketitle

\section{Introduction}\label{sec:introduction}

Despite the recent proliferation of black-box machine learning
techniques such as Neural Networks, symbolic or equation-based models
remain critical in many scientific disciplines owing to their
interpretability and parsimonious nature. The sparse identification of
nonlinear dynamics (SINDy) technique \citep{Brunton2016} has become
one of the most popular techniques to learn the governing dynamical
equation from observed time-series data of the full system state.
Starting from a library of candidate functions that presumably contain
the terms required to describe the system, SINDy's algorithm regresses
the time derivative estimated from the time-series data of the states 
against the nonlinear functions of its
state while promoting sparsity in the resulting learnt parameters. The
resulting parsimonious model was shown to share the same sparsity
pattern as the original model, effectively demonstrating that the SINDy
algorithm can recover governing equations from data given an appropriate
library of nonlinear functions and coordinates for the observed
dynamics. Owing to its simplicity and speed, since its inception, the
method has been applied to many applications across disciplines. It has
also attracted and inspired many improvements, extensions and variants,
ranging from alternative sparsity-promoting regression algorithms (SR3)
\citep{Zheng2019}, extensions to finding PDEs \citep{Rudy2017},
extensions to libraries of rational functions \citep{Kaheman2020},
improvements to noise robustness by ensembling \citep{Fasel2022} to
the incorporation of active learning strategies \citep{Fung2025}.

Despite its utility, SINDy's sparsity promotion process is inherently
empirical, as the underlying STLSQ removes terms by coefficient
magnitude during the thresholding process, preventing the algorithm from
learning potentially multiscale dynamics. Recent works
\citep{Zhang2018,Hirsh2022,Niven2024,Fung2025} have shown that by
adopting a Bayesian perspective on the sparsity-promoting process, one
can not only improve noise robustness in recovering the correct model
but also put the sparsity promotion on a more rigorous foundation. The
key advantage of the Bayesian interpretation of SINDy is that the
principle of Occam's razor that underlies the argument for a
parsimonious model naturally arises when one maximises the Bayesian
evidence or marginalised likelihood \citep{MacKay2003}. Another
approach to improve the robustness of SINDy is to use it in conjunction
with ensembling \citep{Fasel2022}, which, as shown by recent work
\citep{Gao2023}, can also be interpreted as a Bayesian approach.

Although the Bayesian reformulation can somewhat improve noise
robustness, as demonstrated empirically by recent works
\citep{Fasel2022,Niven2024,Fung2025}, large noise remains challenging
to the SINDy framework and many of its variants. This challenge of noise
is twofold. First, the SINDy framework relies on a good approximation of
the time derivatives of time-series data \(\dot{\mathbf{X}}\). Classical
methods to approximate time derivatives, like finite difference, tend to
amplify the measurement noise in time-series data. To alleviate this
noise, \citet{Reinbold2020} and \citet{Messenger2021} indepedently
introduced the weak formulation. It is conceptually similar to applying
a finite element method in the time domain, as the weak form projects
the time series data onto some local trial function space. Irrespective
of how the time derivatives are approximated, whether via finite
difference, weak formulation, or other collocation methods, they can all
be written as a linear transformation of the time series. Therefore, the
propagation of noise is tractable and easily captured by the Bayesian
method. By taking into account the noise variance amplification by the
time-derivative approximation, as demonstrated by \citet{Fung2025}, a
Bayesian formulation can alleviate this noise amplification issue.

Meanwhile, the second challenge posed by large noise is difficult to
address even with the aforementioned Bayesian extension. The challenge
arises from the contamination of the nonlinear dictionary
\(\boldsymbol{\Theta}(\mathbf{X})\), as the adverse effect of noise can
be amplified by highly nonlinear functions. Our previous work
\citep{Fung2025} accounted for this noise contamination to some
extent by propagating the noise through a local linear approximation to
the dictionary. However, the propagation only works if the noise
magnitude is small enough for the approximation. For large noise,
propagation through a nonlinear term will result in a highly
non-Gaussian or even multimodal distribution that evades simple
regression techniques.

Nonetheless, learning an equation from signals with a high
noise-to-signal ratio is not completely impossible. The key insight is
that information regarding the noise-contaminated state at a certain
time is also contained in the neighbouring data points in time. The weak
formulation is one way to extract this information, as it effectively
projects the neighbourhood of a certain time onto a more plausible (and
often smoother) subspace of temporal basis functions. In a follow-up
paper, \citet{Bortz2023} argued that the weak formulation is an
effective way to overcome the error-in-variable problem. Indeed, their
results show that this is a plausible and fast method. However, this
approach inevitably introduces spectral bias, as it effectively assumes
the trajectory follows a certain shape in the local time domain, which
described by a linear combination of the trial basis functions used.
Indeed, both \citet{Messenger2021} and \citet{Fung2025} have
discussed the implication of the weak formulation as a low-pass filter.
Hence, the use of weak formulation can be interpreted as a form of
inductive bias that regularises the learning problem. 
It encodes the user's priorknowledge of the frequency of the true signal. 
If the data is sampled at a much higher frequency than the signal, 
the weak formulation is an effective way to filter the noise away from the signal.

A less biased approach is to use the model itself to extract information
from neighbouring points. After all, the model, in contrast to the weak
form trial basis, is the true statement that describes the relationship
between neighbouring points. The key task in this approach is,
therefore, to leverage the learnt model to estimate the original states
\(\mathbf{X}\) from information provided by neighbouring data points.
However, the model is not known \emph{a priori}. Hence, the challenge is
to simultaneously denoise the data and learn the equation in a single
optimisation. Several recently proposed methods, including
modified-SINDy \citep{Kaheman2022}, DAHSI \citep{Ribera2022},
DySMHO \citep{Lejarza2022} and SIDDs \citep{Hokanson2023}, have
attempted this approach with various optimisation and modelling
strategies. Modified-SINDy \citep{Kaheman2022} put the numerical
integration of the ODE inside an optimisation that optimises the model
parameter and promotes sparsity while minimising the discrepancy between
the trajectorial data and the integrated path. DAHSI
\citep{Ribera2022}, DySMHO \citep{Lejarza2022}, and SIDDs
\citep{Hokanson2023} used a similar strategy but discretised the
equation by some collocation or finite difference method first before
optimisation, thereby preventing the use of expensive ODE solvers within
the optimisation. These works have claimed improvements in noise
robustness and accuracy in the parameter estimate compared to previous
linear regression-based methods \citep{Fasel2022}.

However, with the exception of DAHSI \citep{Ribera2022}, a common
feature in the aforementioned works
\citep{Kaheman2022,Lejarza2022,Hokanson2023} is that the differential
equation acts as a hard constraint that needs to be satisfied exactly
during optimisation. For modified-SINDy \citep{Kaheman2022}, this
enforcement comes in the form of integrating the ODE with classic
solvers. For DySMHO \citep{Lejarza2022} and SIDDs
\citep{Hokanson2023}, both of which employ a discretise-then-optimise
approach, the enforcement of the equation comes in the form of a
constrained optimisation, in which the denoised trajectories are
constrained to follow a set of parameterised equations, the parameter
and form of which are to be ``learnt'' via sparsity-promoting
optimisation. The ``hard'' constraint in these optimisations creates
stiffness in the optimisation that prevents assimilation of data longer
than the Lyapunov timescale when the system is chaotic. To relax this
stiffness, all of the aforementioned works have implicitly applied some
form of relaxation empirically, which we shall more explicitly point out
in \cref{sec:stochastic_relax}.

In this work, we introduce ODR-BINDy, which also exploits the noise
robustness of the denoising strategy. In many ways, ODR-BINDy is similar
to modified-SINDy, DySMHO, DAHSI and SIDDs in its ability to denoise and
learn simultaneously. Much like DySMHO, DAHSI, and SIDDs, ODR-BINDY also
employs a discretize-then-optimize strategy. However, there are a few
key differences where ODR-BINDy stands out. 1. ODR-BINDy, like our
previous BINDy, uses Bayesian evidence to select models, which has been
shown to be more rigorous and robust. 2. ODR-BINDy employs a more
traditional Newton's method to optimise the underlying nonlinear least
square problem, which, as we shall demonstrate later, converges much
quicker with a good initial guess, which one can find by ensemble linear regression.
3. Most importantly, ODR-BINDy relaxes the fulfilment of the
differential equation by imposing it as a soft constraint, making it
more robust against discretisation error or stochastic noise.

\section{Formulation}\label{formulation}

We start by introducing the formulation of SINDy or a typical SINDy
variant. Firstly, it assumes that the data is driven by an underlying
(autonomous) dynamical system that can be written in the form
\begin{equation}\protect\phantomsection\label{eq:dyn_sys}{
\dot{\mathbf{x}} = \mathbf{f}(\mathbf{x}(t),\boldsymbol\Xi) =  \boldsymbol{\Theta}(\mathbf{x}(t))\boldsymbol\Xi,
}\end{equation} where
\(\mathbf{x}(t) = [x_1(t) \; x_2(t) \; \dots \; x_j \; \dots \; x_D(t)] \in \mathbb{R}^{D}\)
are the states of a \(D\)-dimensional dynamical system,
\(\boldsymbol{\Theta}(\mathbf{x}(t)) \in \mathbb{R}^M\) a dictionary of
\(M\) nonlinear functions of the states, and
\(\boldsymbol{\Xi} \in \mathbb{R}^{M\times D}\) the parameters that
parameterise the dynamical system. SINDy then assumes a parsimonious
description of the function \(\mathbf{f}\) would imply sparsity in the
parameter matrix \(\boldsymbol{\Xi}\in \mathbb{R}^{M\times D}\), given
the right coordinate for the states and the right dictionary is chosen
to describe the system. Therefore, by approximating \(\dot{\mathbf{x}}\)
with the appropriate method and regressing it against
\(\boldsymbol{\Theta}(\mathbf{x}(t))\) while promoting the correct
sparsity, one can effectively learn the functional form of
\(\mathbf{f}\). It should be noted that \(\mathbf{f}\) does not
necessarily always have the form described in \cref{eq:dyn_sys}, i.e.,
linearly summed nonlinear functions of \(\mathbf{x}\). In particular,
DySMHO \citep{Lejarza2022} and DAHSI \citep{Ribera2022} have shown
how the framework can be extended beyond \cref{eq:dyn_sys}.

In this work, we will assume the data comes from \textbf{full and unbiased observation of all the states}, meaning that we can directly measure all states \(\hat{\mathbf{x}}(t_i)\)
at time \(t= t_1, t_2 , \dots t_i \dots t_{\tilde{N}}\), with
some unbiased measurement noise \(\boldsymbol{\epsilon}_x\), such that
\begin{equation}\protect\phantomsection\label{eq:noise_model}{
\hat{{X}}_{ij} =\hat{{x}}_j(t_i) = {x}_{j}(t_i)+\epsilon_{x,ij} = {X}_{ij} + \epsilon_{x,ij}.
}\end{equation} 
For simplicity, we assume the sampling is done at a
fixed time interval, although this is not necessarily a restriction for
the methodology. 
However, SINDy-like methodologies, as well as this work, necessitate that the sampling be frequent enough. 
For example, in the context of chaotic systems, the sampling should be more frequent than the inverse of Lyapunov time.
We will also assume that the noise
\({\boldsymbol{\epsilon}}_x \in \mathbb{R}^{\tilde{N} \times D}\)
consists of \(\tilde{N} \times D\) i.i.d. random variables with Gaussian
distributions, with zero means (i.e. unbiased) and variance
\(\boldsymbol\sigma_x^2 \in \mathbb{R}^{\tilde{N} \times D}\).
Extensions to other noise models are possible by substituting the loss
function in the next step with the corresponding energy function
(negative log of the probability distribution representing the noise
model). For simplicity, here we assume Gaussian noises.

Given the discrete sampling, the dictionary
\(\boldsymbol{\Theta}(\mathbf{x}(t))\) is rewritten as a matrix
\(\boldsymbol\Theta(\mathbf{X}) \in \mathbb{R}^{\tilde{N} \times D}\),
where each row represents each time point, and each column represents
each nonlinear function. For example, if the dictionary is assumed to be
polynomials of the states \(\mathbf{x}(t)=[x_1(t),x_2(t),x_3(t)]\) of a
three dimensional system (i.e.~\(D=3\)), we can write down the
dictionary as \[
\boldsymbol\Theta(\mathbf{X})=
\begin{bmatrix}
1 & X_{11} & X_{12} & X_{13} & X_{11}^2 & X_{11}X_{12} & X_{11}X_{13} & X_{12}^2 & X_{12}X_{13} & X_{13}^2 & \dots \\
1 & X_{21} & X_{22} & X_{23} & X_{21}^2 & X_{21}X_{22} & X_{21}X_{23} & X_{22}^2 & X_{22}X_{23} & X_{23}^2 & \dots  \\
1 & X_{31} & X_{32} & X_{33} & X_{31}^2 & X_{31}X_{32} & X_{31}X_{33} & X_{32}^2 & X_{32}X_{33} & X_{33}^2 & \dots  \\
\vdots & \vdots & \vdots & \vdots & \vdots & \vdots & \vdots & \vdots & \vdots & \vdots & \ddots 
\end{bmatrix}
\] up to some order that also determines \(M\). To approximate the time
derivative from the time-series data \(\mathbf{X}\), we assume there
exist some linear operators in the form of matrices
\(L_I \in \mathbb{R}^{N \times\tilde{N}}\) and
\(L_{\partial t}  \in \mathbb{R}^{N \times\tilde{N}}\) such that
\begin{equation}\protect\phantomsection\label{eq:Collocation}{
L_{I} \dot{\mathbf{X}}_i \approx L_{\partial t} \mathbf{X}, \quad \mbox{and therefore}  \quad L_{\partial t} \mathbf{X} \approx  L_{I} \mathbf{f}(\mathbf{X})
}\end{equation} Both finite difference and weak formulation, as well as
other collocation-based methods that discretise the time domain, can be
written in this form. Moreover, \(L_I\) and \(L_{\partial t}\) are
usually sparse or banded, accelerating the computation in later steps. A
detailed discussion on these operators can be found in
\citep{Fung2025}.

\begin{figure}
\includegraphics[width=1.0\linewidth,keepaspectratio]{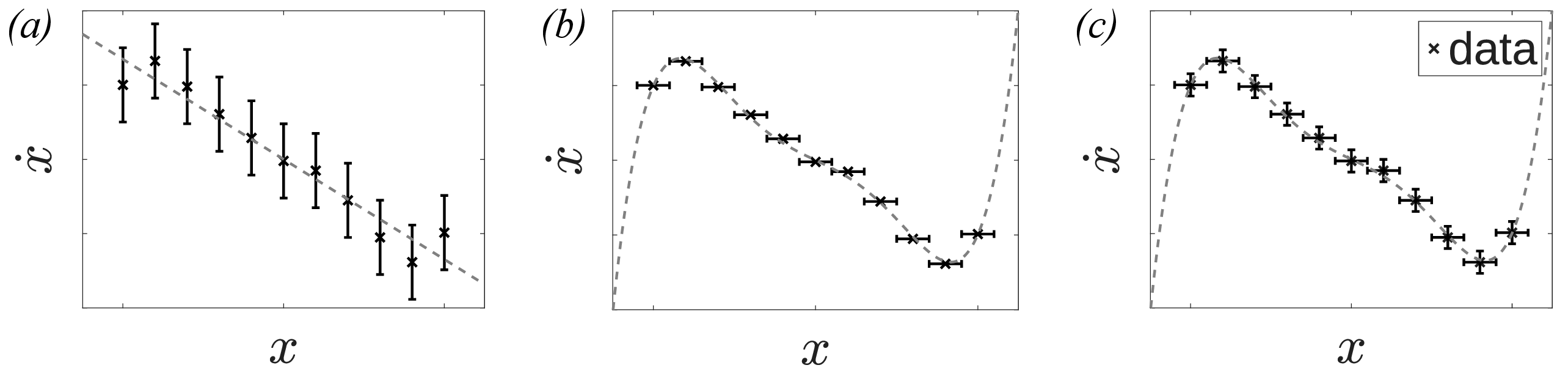}

\caption[{Visualisation of noise type by the direction of errorbars.
\((a)\) Linear regression-based methods, like SINDy, assumed that the
noise is in the \(\dot{\mathbf{x}}\) direction only. \((b)\) Nonlinear
optimisation-based methods, like modified-SINDy, DySMHO and SIDDs, uses
hard constrain to satisfy the equation, equivalent to assuming no noise
in the \(\dot{\mathbf{x}}\) direction. Hence, noise is in the
\({\mathbf{x}}\) direction only. \((c)\) Soft-constrained nonlinear
optimisation-based methods, like DAHSI and ODR-BINDy, assumed noise in
both the \({\mathbf{x}}\) and \(\dot{\mathbf{x}}\) direction. Regression
with noise in the regressors is also called error-in-variable analysis.
One technique to perform this regression is the orthogonal distance
regression (ODR) \citep{Boggs1987}.}]{Visualisation of noise type by
the direction of errorbars. \((a)\) Linear regression-based methods,
like SINDy, assumed that the noise is in the \(\dot{\mathbf{x}}\)
direction only. \((b)\) Nonlinear optimisation-based methods, like
modified-SINDy, DySMHO and SIDDs, uses hard constrain to satisfy the
equation, equivalent to assuming no noise in the \(\dot{\mathbf{x}}\)
direction. Hence, noise is in the \({\mathbf{x}}\) direction only.
\((c)\) Soft-constrained nonlinear optimisation-based methods, like
DAHSI and ODR-BINDy, assumed noise in both the \({\mathbf{x}}\) and
\(\dot{\mathbf{x}}\) direction. Regression with noise in the regressors
is also called error-in-variable analysis. One technique to perform this
regression is the orthogonal distance regression (ODR)
\citep{Boggs1987}.}

\label{fig:errorbars}

\end{figure}

Here, ODR-BINDy, along with the aforementioned works
\citep{Kaheman2022,Ribera2022,Lejarza2022,Hokanson2023}, departs from
classical linear regression-based SINDy variants (including BINDy
\citep{Fung2025}). Classical SINDy-like approaches presumed that all noise
resides in the regressand, specifically on the left-hand side of the second equation of \cref{eq:Collocation}. 
In particular, \citep{Fung2025} had spent
considerable effort approximating how noise in \(\mathbf{X}\) propagates
through the nonlinear dictionary \(\boldsymbol{\Theta}(\mathbf{X})\)
such that one can effectively treat the noise as solely present the
regressand and not in the regressors. \Cref{fig:errorbars}$(a)$ illustrates this
approach in a simple 1D case, in which the noise is expressed as error
bars in the \(\dot{\mathbf{x}}\) direction. The advantage of this approach
lies in the assumption that the effective noise exists just in the
regressand, allowing us to utilise basic linear
regression to reduce the least squares. The drawback, however, is that
the effective noise may not conform to a Gaussian distribution,
as presumed by the least square method. This is
especially true when large noise is propagating through highly nonlinear
functions, thereby limiting the method's robustness against large noise.

However, if we stay true to the noise model described in
\cref{eq:noise_model}, the noise should have been in the regressand instead of the regressor. Visually, we can represent this
regression as a minimisation in the regressor error as shown in
\cref{fig:errorbars}$(b)$, which is in contrast to the minimisation of regressand
error by linear regression.
If we also assume a Gaussian prior on the parameter, the justification of which will be discussed in \cref{sec:bayesian-model-selection} (also see \citep{Fung2025}), then the loss function arise from the log-posterior
distribution should be 
\begin{equation}\protect\phantomsection\label{eq:Nonlinear-loss}{
\mathcal{L}(\mathbf{X},\boldsymbol{\Xi}) = \frac{1}{2\sigma^2_x}||\hat{\mathbf{X}}-\mathbf{X}(\boldsymbol{\Xi})||_F^2 + \frac{1}{2 \sigma_p^2}||\boldsymbol{\Xi}||^2_F,
}\end{equation} where \(\mathbf{X}(\boldsymbol{\Xi})\) is constrained by
the discretised differential equation
\begin{equation}\protect\phantomsection\label{eq:constrain}{
\quad L_{\partial t} \mathbf{X} -  L_{I} \boldsymbol{\Theta}(\mathbf{X})\boldsymbol\Xi = 0.
}\end{equation} 
In this case, we can avoid the inconveniences of
approximating noise propagation through the nonlinear functions in
\(\boldsymbol{\Theta}(\mathbf{X})\). This approach is analogous to how linear
regression is extended to nonlinear regression in a typical
error-in-variables (EIV) analysis, and form the basis for the standard
way to perform parameter estimation by data assimilation
\citep{Ljung2010,Asch2016}. Indeed, minimising
\cref{eq:Nonlinear-loss,eq:constrain} is akin to the strong constraint
four-dimensional variational (4D-Var) assimilation technique
\citep[chap.~2.4.3.2]{Asch2016} . DySMHO, SIDDs and modified-SINDy are
also solving this optimisation problem (or its equivalence). We also
note that the loss in \cref{eq:Nonlinear-loss} can be easily modified to
account for noise models other than Gaussian. Although this nonlinear
optimisation has no guarantee for convexity, it is not a major
constraint provided that we have a good initial guess. This initial
guess can be from the data itself or some pre-processed form of it that
has already smoothed some of the data (see \cref{sec:init_guess}).

\subsection{The importance of stochastic
relaxation}\label{sec:stochastic_relax}

Now, besides a slightly different prior and sparsity promotion methodology,
\cref{eq:Nonlinear-loss,eq:constrain} are effectively the optimisation
problem underpins DySMHO and SIDDs. Modified-SINDy is also solving an
equivalent problem, but it enforced the constraint by evaluating the ODE
with a solver rather than use collocation in the time domain. Here, our
ODR-BINDy approach and DAHSI differ from DySMHO, modified-SINDy, and
SIDDs: \textbf{Rather than strictly enforcing
\cref{eq:constrain}, we define the residue to \cref{eq:constrain} as a
small noise} \[
\boldsymbol\eta =  L_{\partial t} \mathbf{X} -  L_{I} \boldsymbol{\Theta}(\mathbf{X})\boldsymbol\Xi .
\] The rationale for this noise \(\boldsymbol\eta\) is as follows:
Real-life data seldom adhere precisely to the dynamics outlined by
\cref{eq:dyn_sys}. Instead, there is always small stochasticity that
results in the state deviating somewhat from the anticipated dynamics
described by \cref{eq:dyn_sys}. Even with synthetic data where
stochastic noise is ostensibly absent, numerically solving
\cref{eq:dyn_sys} will inevitably introduce truncation error. The
collocation method that gives rise to \(L_{\partial t}\) and \(L_{I}\)
also introduces truncation error. In other words, \cref{eq:constrain}
can never be satisfied exactly (except when the identical discretisation
scheme is employed for both data generation and assimilation, which is
rarely the case in practice), and the noise \(\boldsymbol\eta\)
represents any discrepancy between the learnt model and the dynamic that generated the data.

Now, in systems without any positive Lyapunov exponent, a small
\(\boldsymbol\eta\) will not prevent methods like DySMHO, SIDDs and
modified-SINDy from enforcing \cref{eq:constrain} to the best accuracy
possible when solving the inverse problem through optimising
\cref{eq:Nonlinear-loss} . However, for chaotic systems or any system
with a positive Lyapunov exponent, the small error \(\boldsymbol\eta\)
will accumulate in time, preventing \(\mathbf{X}\) from following the
data \(\hat{\mathbf{X}}\) for a time longer than the Lyapunov time
scale. In other words, if \cref{eq:constrain} is enforced exactly
without relaxation, i.e., as a ``hard'' constraint, a small mismatch
between how the data is generated \cref{eq:dyn_sys} and how the data is
modelled \cref{eq:constrain} or slight stochasticity in the real-life
data will prevent \(\mathbf{X}\) from following the data
\(\hat{\mathbf{X}}\) longer than the Lyapunov time scale.

\begin{figure}
\centering
\includegraphics[width=0.49\linewidth,height=\textheight,keepaspectratio]{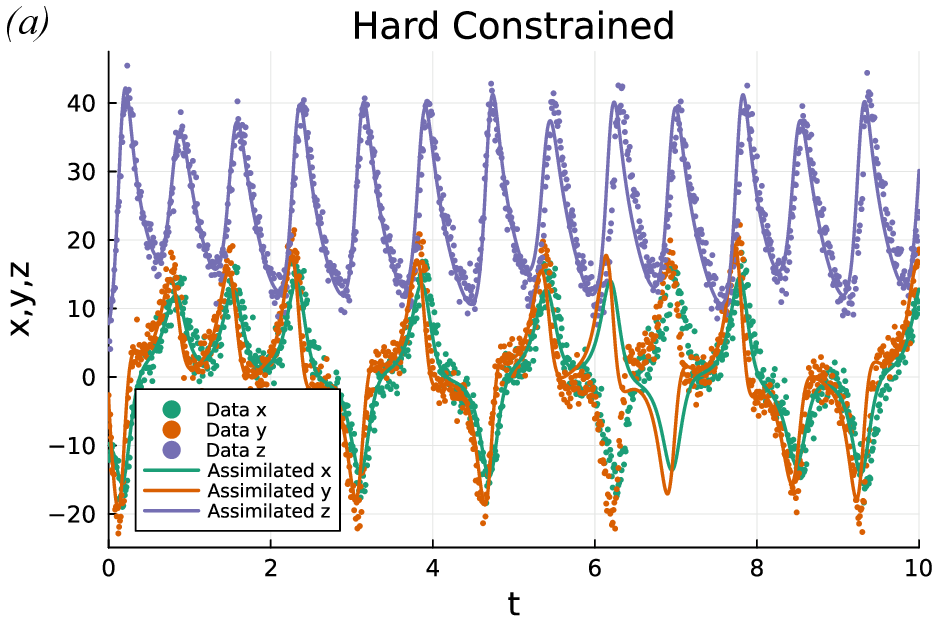}\label{fig:LorenzAssim_Hard}
\includegraphics[width=0.49\linewidth,height=\textheight,keepaspectratio]{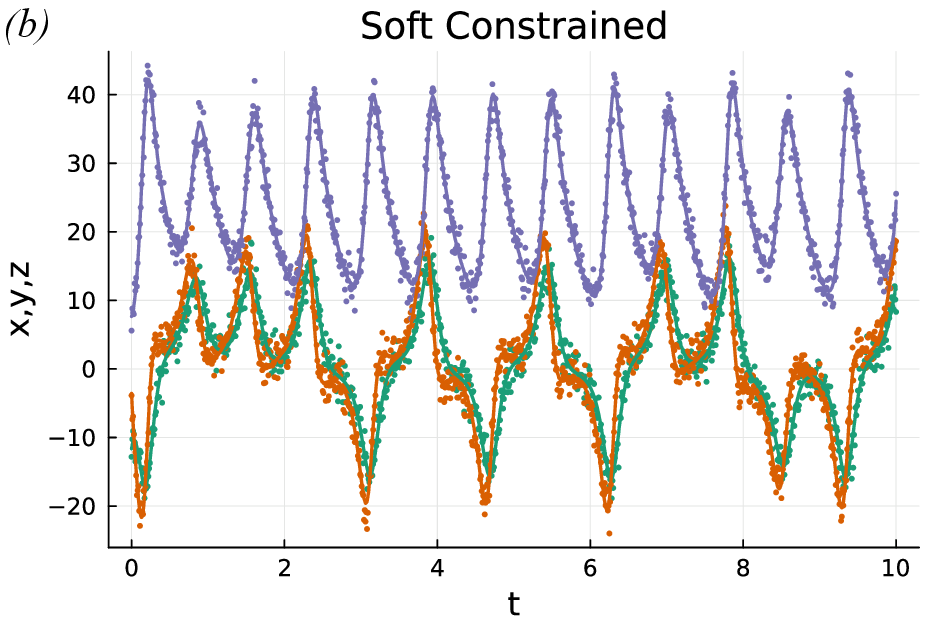}\label{fig:LorenzAssim_Soft}\\
\includegraphics[width=0.80\linewidth,height=\textheight,keepaspectratio]{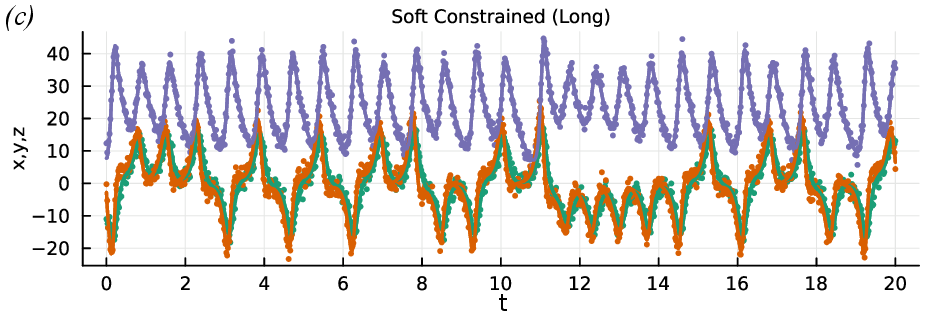}\label{fig:LorenzAssim_Soft_Long}

\caption{Data assimilation of Lorenz63 with \((a)\)
optimisation upon hard constrained ODE solver and \((b)\) with soft
constrained loss function. Data are sampled at \(\Delta t =0.01\) with
noise \(\sigma_x=1.6\). Notice how the assimilated trajectory diverge at
\(t \sim 7\) in \((a)\). This is despite our best effort to optimise the
parameter and the initial condition. This is because the constraint
\cref{eq:constrain} do not account for the any numerical error that will
lead to divergence longer than Lyapunov time scale. In contrast, the
soft constrain \cref{eq:loss_soft} use in \((b)\) relax the problem such
that the assimilated data can always follow the data in the long time limit, even beyond Lyapunov time scale. However, the data must be sampled more frequently than Lyapunov time. \((c)\) Assimilation of a longer trajectory using the same soft constraint as \((b)\) to demonstrate that the method indeed works for any arbitrarily long series of data.}

\label{fig:LorenzAssim}

\end{figure}

\Cref{fig:LorenzAssim} illustrates this difficult in a prototypical
problem of parameter estimation in Lorenz63, whereby data is
assimilated to estimate the parameters of know equations of Lorenz63 by
optimising \cref{eq:Nonlinear-loss} under the constrain
\cref{eq:constrain}. It is evident that without relaxing the hard
constraint, the model cannot fully adhere to the data.

We should note that this is not a novel discovery. The challenge of
assimilating data beyond the Lyapunov time scale is well established
\citep{Bard1974,Ramsay2007,Asch2016}. However, in the context of
model discovery, DAHSI is the sole notable study known to us that
explicitly incorporates the model error \(\boldsymbol\eta\) as a soft
constraint rather than a hard constraint. Instead, most studies
\citep{Kaheman2022,Lejarza2022,Hokanson2023} address the issue
through empirical relaxation during training. For instance, SIDDs
\citep{Hokanson2023} implicitly relaxed the constraint
\cref{eq:constrain} during the training, enforcing \cref{eq:constrain}
only at termination. Modified-SINDy empirically penalises losses that
are not temporally local through the \(\omega\) prefactor in eq. (11-12)
of \citep{Kaheman2022}. Meanwhile, DySMHO \citep{Lejarza2022}
limited the time horizon to be smaller than the Lyapunov time scale.
While these empirical relaxations somewhat attenuate the issue, they
function more as a hotfix than a theoretically informed relaxation and
may hinder the method's capacity to maximise learning from the given
data.

In this study, we propose that rather than relaxing the issue by other means, we should treat the error \(\boldsymbol\eta\) arising from numerical
truncation or small stochasticity properly as part of the loss. 
With the risk of slight over-simplification, here we empirically model \(\boldsymbol\eta\) as an independent and identically distributed noise
\(\boldsymbol\eta \sim \mathcal{N}(0, \sigma_{\partial t}^2)\),
characterised by a Gaussian distribution with variance
\(\sigma_{\partial t}^2\), which quantifies the magnitude of such error.
The use of Gaussian distribution as an approximation mostly serves the
purpose of simplicity, and it may readily be substituted with
alternative noise models. With the injection of this noise, the model
equation becomes analogous to a random ODE or SDE; thus, we shall
designate this technique as \textbf{stochastic relaxation}. 
The phrase also highlights the fact that although the system to be learnt is deterministic, to assimilate data beyond Lyapunov time, one must treat it as if it has some stochasticity in order to overcome the system's sensitivity to small perturbations arising from numerics. 
While this modification may seem inconsequential at first, its significance cannot
be understated. The fact that we allow the learning algorithm to adjust
the integration of \cref{eq:constrain}, however small the adjustment is,
allows the integrated trajectory from \cref{eq:constrain} to always
follow the data, even beyond Lyapunov timescale. This helps the
algorithm maximise the information gain from a dataset by assimilating
long trajectories in a single optimisation, which is in contrast to the
moving horizon strategy of DySMHO and the empirical penalisation of
modified-SINDy. In practice, stochastic relaxation reformulate the
problem into an unconstrained optimisation for the loss
\begin{equation}\protect\phantomsection\label{eq:loss_soft}{
\mathcal{L}(\mathbf{X},\boldsymbol{\Xi}) = \underbrace{\frac{1}{2\sigma^2_x}||\hat{\mathbf{X}}-\mathbf{X}(\boldsymbol{\Xi})||_F^2}_{\text{Data Loss}} + \underbrace{\frac{1}{2 \sigma^2_{\partial t}} || L_{\partial t} \mathbf{X} -  L_{I} \boldsymbol{\Theta}(\mathbf{X})\boldsymbol\Xi ||^2_F}_{\text{Model Loss}} + \underbrace{\frac{1}{2 \sigma_p^2}||\boldsymbol{\Xi}||^2_F}_{\text{Prior}}.
}\end{equation}

In other words, we have essentially regarded \cref{eq:constrain} as a
``soft'' constraint rather than a ``hard'' constraint.
We designate the first term as data
loss, next term as model loss, and the last term as the prior. Each term
is scaled by the corresponding noise and prior variance. Visually,
\cref{fig:errorbars}$(c)$ depicts the noise variances as error bars: the error
bar in the \(\mathbf{X}\) direction depicts \(\sigma_x\), and the error
bar in the \(\mathbf{\dot{X}}\) direction depicts
\(\sigma_{\partial t}\). Since the noise is in both the \(\mathbf{X}\)
direction and \(\mathbf{\dot{X}}\) direction, the overall optimisation
is analogous to Orthogonal Distance Regression (ODR). Hence, we shall
directly apply the same regression technique for ODR in this work.

Here, we note that optimising
\cref{eq:loss_soft} is akin to the weak constraint four-dimensional variational (4D-Var) technique in data assimilation 
\citep[chap.~2.4.3.2]{Asch2016}. It is also equivalent to the loss function used in Physics-Informed Neural Network (PINN). Indeed, there is a deeper theoretical connection between 4D-Var data assimilation, PINN and \cref{eq:loss_soft}. For the sake of brevity, we will address this connection in a future publication.

Notably, the package DAHSI also shares the same loss function. 
However, our approach to managing the hyperparameters and the philosophy behind the derivation of the loss function is different. 
In DAHSI's variational annealing method, the scaling
pre-factor of the model loss is gradually adjusted throughout
optimisation. In contrast, we fix each hyperparameter in accordance
their mathematical origin: each pre-factor scaling the loss functions is
the inverse of the corresponding noise and prior variance, fixed
according to the user's knowledge or assumptions regarding measurement
noise, stochastic noise or truncation error, and the prior. More
discussion on how hyperparameters are chosen will be discussed in
\cref{sec:hyperparameters}.

\subsection{Bayesian model selection}\label{sec:bayesian-model-selection}

After the loss function is established, the subsequent step is
implementing the model selection process. To this end, we utilise the
same sparsity inducing hierarchical prior and Greedy technique as outlined in \citep{Fung2025}, wherein we
calculate, with a suitably chosen Gaussian prior, the Bayesian evidence via Laplace Approximation for each model
and sequentially discard terms with the least evidence. The justification to use a zero-mean the Gaussian prior as an approximation to the true sparsity-inducing hierarchical prior in the context of maximizing Bayesian evidence can be found in \citep{Tipping2003,Faul2020} and \S2$(b)$ of \citep{Fung2025}. To expedite
calculations and guarantee that the parameter variance remains positive
definite, we also utilise the Gauss-Newton approximation for the
evidence, as this approximation appears to have minimal impact on the
model selection process (see \cref{sec:GaussNewton}).

The benefits of choosing a model based on Bayesian evidence, as opposed
to thresholding or other empirical methods, have been extensively
examined in \citep{Fung2025}. We succinctly summarise the findings
herein. The Bayesian evidence measures the likelihood of the data
conditional on the hypothesised model. Assuming that all potential models (i.e.,
combinations of candidate functions in the library) are equally
probable, the evidence measures the extent to which a model explains the
data. It also naturally encourage more parsimonious models
\citep{MacKay1992}. Consequently, it serves as a mathematically
rigorous metric for comparing various models and promoting sparsity
within the model \citep{MacKay2003}. Identifying the model that
optimises evidence is also referred to as type-II maximum likelihood
\citep{Tipping2003}. Unlike the empirical thresholding procedure of
STLS that underpins SINDy and its numerous variants, which often
eliminates significant terms with small coefficients, the Bayesian
method does not discards coefficients according to their magnitude.
Instead, it eliminates coefficients are too sensitive to minor
perturbations, which is shown to be a more rigorous approach
\citep{MacKay1992,Tipping2003,MacKay2003,Niven2024}. In this way, it
is similarly analogous to the coefficients of variation utilised in
\citep{Lejarza2022}. Works by \citep{Fung2025,Niven2024} have
demonstrated that, in the context of linear regression-based model
discovery, employing Bayesian evidence to enhance sparsity promotion
significantly increases the likelihood of accurately selecting the
optimal model in sparse and noisy datasets. This study will illustrate
similar results in the context of the nonlinear optimisation approach
error-in-variable regression, by introducing the Orthongal Distance
Regression-based Bayesian Identification of Nonlinear Dynamics, a.k.a.
ODR-BINDy, and comparing it with other STLS-based methodologies.

\subsection{The ODR-BINDy algorithm and finding a good initial
guess}\label{sec:init_guess}

The workflow of the ODR-BINDy is summarised as follow. First, we conduct
a nonlinear optimisation to minimise \cref{eq:loss_soft} by adjusting
\((\mathbf{X},\boldsymbol{\Xi})\) while incorporating all library terms.
The data \(\hat{\mathbf{X}}\) functions as the initial guess for
\(\mathbf{X}\), whereas multiple initial guesses for
\(\boldsymbol{\Xi}\) is provided by bootstrap linear regression. Our
experience shows that the bootstrap strategy produce relatively good
initial guess, which is consistent with previous literature
\citep{Fasel2022}. The randomness in bootstrap also gives multiple
initial guess, allowing the use of multi-start as a global optimisation
algorithm. Subsequently, in accordance with the Greedy algorithm, we
evaluate all models that possess one fewer term than the preceding
iteration by executing the same ODR for each trial model and
calculating the Bayesian evidence for each model. Despite the apparent
computational expense of executing many nonlinear optimisations, this
phase, which constitutes the majority of the algorithm's duration, can
be substantially accelerated by the following techniques:

\begin{enumerate}
\item   Supplying the analytical gradient to the optimiser can markedly
  accelerate computation. It is possible because the library function
  and its derivatives can be easily obtained. Furthermore, given that
  \(L_{\partial t}\) and \(L_I\) often appear as sparse matrices, the
  gradient with respect to \(\mathbf{X}\) is also sparse. This sparsity
  can substantially accelerate optimisation based on Newton's method.
\item   In each successive trial of new models with one less term, the outcome
  \((\mathbf{X},\boldsymbol{\Xi})\) from the preceding iteration with
  more terms is typically a reliable initial guess for the next
  nonlinear optimisation. This allows successive optimisation to be sped
  up significantly. Indeed, convergence using Newton's approach is often
  achieved within one hundred iterations if the initial guess is close
  enough.
\item   Conversely, if the trial model lacks the required term, the
  optimisation frequently fails to converge. This prompts us to restrict
  the maximum number of steps during model trials and to retain the term
  whose elimination would result in non-convergence.
\item   The calculation of Bayesian evidence relies on the Laplace
  approximation, which only requires the parameter covariance at the
  minimum loss point. This computation can be further accelerated with
  the application of Gauss-Newton estimation, which also ensures the
  positive definiteness of the covariance. The full formulation of the
  evidence and Hessian is provided in \cref{sec:EvidenceFormula}, and
  the impact of Gauss-Newton estimation in place of full Hessian
  calculation can be found in \cref{sec:GaussNewton}.
\item   The computation of each trial model relies solely on the preceding run
  for an optimal initial guess and is independent of one another.
  Consequently, it can be further parallelised on a multi-node or
  multi-core CPU.
\end{enumerate}

Upon evaluating all models with one fewer term, we select the model with
the most evidence and proceed to the subsequent term elimination
following the Greedy technique. The algorithm terminates and selects the
global maximum evidence model when it observes a successive decline in
evidence following the elimination of several terms. The procedure
consequently produces the trained sparse model along with its parameters
and covariance, in addition to the associated (log-)Bayesian evidence
that measures the model's quality.

\subsubsection{Choice of nonlinear optimization algorithm and
computational
speed}\label{choice-of-nonlinear-optimization-algorithm-and-computational-speed}

As mentioned before, optimising \cref{eq:loss_soft} is effectively a
ridge Orthogonal Distance Regression \citep{Boggs1987}. Given the
local convexity of each trial model using good starting guesses from
previous model iteration and the sparse analytical gradient, in this
work we chose the \texttt{lsqnonlin} routine from \texttt{MATLAB} as the
underlying optimisation algorithm. This algorithm is based on Newton's
method with a trust region, specifically optimised for nonlinear least
squares problems. Since \(L_{\partial t}\) and \(L_I\) are sparse banded
matrices, the computational complexity of the optimisation is
\(\mathcal{O}(N Db^2)\), where \(b\) is the width of the banded matrix.
As mentioned, to encourage convergence to the global minimum, for the
first iteration where the full library is used, we also employ a
multi-start global optimisation strategy, using the bootstrapped
estimate of \(\boldsymbol{\Xi}\) serves as the probabilistic starting
point.

To select a parsimonious model, the \texttt{lsqnonlin} routine is
wrapped inside a greedy algorithm that tests the removal of each term,
as described above. Therefore, the overall computational complexity
scales with \(\mathcal{O}(M^2 N Db^2)\), in which \(M\) is the number of
terms in the library. Despite the polynomial scaling, the rapid
convergence and the use of stochastic relaxation have provided such a
speedup in computation that the overall learning speed of our proposed
algorithm remains faster than other existing methods. For example, to
learn the Lorenz63 system with 1000 data points and 20\% signal-to-noise
ratio from a 2nd order polynomial library, the method converges to the
correct solution in \textless90s on an Apple M4 Macbook Air, and
\textless40s with 10-cores parallelisation, which is a substantial
improvement over the \textgreater300s performance of DySHMO (with
\texttt{CONOPT}, a proprietary C-based nonlinear optimiser) and
\textgreater130s modified-SINDy (with \texttt{TensorFlow}). DAHSI (with
\texttt{IPOPT} and HSL acceleration, an open-source C-based nonlinear
optimiser) was the only solver that has a faster performance
(\textasciitilde12s) excluding the thresholding parameter sweep loop and
the time to compile the generated script (\textasciitilde40 s), but the
solver also performs poorly in recovering the correct equation without
the hyperparameter sweep (see \cref{sec:Appendix}). It is likely that if
the ODR-BINDy algorithm is implemented in a low-level language such as
C, it will achieve similar performance.

\subsection{Hyperparameter tuning}\label{sec:hyperparameters}

In the present Bayesian framework, the hyperparameters governing the
sparsity promotion process are the measurement variance \(\sigma^2_x\),
model error variance \(\sigma^2_{\partial t}\), and prior
\(\sigma_p^2\). The inverses of these hyperparameters also serve as
prefactors for their respective terms in the loss function. The
measurement noise \(\sigma^2_x\) is typically provided in real-world
datasets (as a measurement error) or artificially introduced in
synthetic datasets. To ensure proper model selection within the Bayesian
framework, it is optimal to utilise its designated value. Meanwhile, the
model error variance \(\sigma^2_{\partial t}\) originates from the
truncation error associated with the discretisation of the temporal
domain or from mild stochasticity in the data. Discretisation error can
be assessed based on the discretisation scheme and sample frequency;
however, for the majority of real-world data, the minor stochasticity
present is challenging to measure \emph{a priori}. In this instance, it
is advisable to conduct an additional hyperparameter optimisation for
\(\sigma^2_{\partial t}\) to maximise the evidence
\citep{MacKay1992}.

Lastly, the prior variance \(\sigma_p^2\), which represents prior
knowledge regarding the magnitude of the learnt coefficients, should be
sensibly large, i.e., larger in order of magnitude than the expected
values of the coefficients such that the prior is weakly informing. As
discussed in \citep{Fung2025}, the learning outcome is not sensitive
to \(\sigma_p\) as long as it is in the appropriate order of magnitude,
and its effect diminishes with an increasing number of data points. It
is advisable to perform another hyperparameter optimisation or
hyperparameter sweep for \(\sigma^2_{p}\) such that the evidence is
maximised \citep{MacKay1992}; but, in practice, a reasonable value
for \(\sigma_p^2\) often produces outcomes very close to the optimised
ones.

\section{Examples}\label{sec:examples}

In the following section, we compare the performance of ODR-BINDy
against the following existing packages that employ the nonlinear
optimisation approach (i.e., optimising for both the parameter
\(\boldsymbol{\Xi}\) and the estimated denoised state \(\mathbf{X}\)),
as well as Ensemble-SINDy \citep{Fasel2022}, which exemplifies the
linear regression-based approach. We will use synthetic data from known
dynamical systems, specifically the Lorenz63, R{\"o}ssler and Van Der Pol
systems, as case studies to assess their efficacy in recovering
information regarding the ground truth. Since the primary objective of
this work is to reconstruct the governing equations from data, the
principal metrics for assessment will be the success rates in recovering
the correct sparsity pattern from a given polynomial dictionary. The
respective model parameter errors are also calculated and shown in
\cref{sec:param_error}. While we acknowledge the existance of
alternative measures (see \citep[sec.~3.1]{Kaheman2022}) for each
package's efficacy, such as the forecast prediction error and KL
divergence from the time-averaged statistics, for the intent and purpose
of this work, we will focus on comparing the success rates.

To our best effort, we have tried to tune the hyperparameters of each
package according to the package's documentation to maximise their
performance. If the example exists in the package's repo, we always try
to follow the default settings from the package, assuming they are tuned
to maximise performance. However, we should note that we did not perform
a full hyperparameter sweep, and the settings and hyperparameters used
may have room for improvement.

In trying to compare against existing packages, we have also noted the
following for some of the packages:

\begin{itemize}
\item   SIDDs: Despite our utmost efforts, we are unable to execute the
  package from their open-source repository due to absent dependencies;
  nonetheless, it is noteworthy that they cited better performance over
  modified-SINDy \citep{Hokanson2023}. The appendix
  \cref{sec:Appendix} illustrates the comparison of SIDDs with ODR-BINDy
  based on our best approximation to the performance they've cited.
\item   DySMHO: We see that DySMHO necessitates significantly more data to
  achieve comparable noise robustness to the other approaches; hence, we
  have provided DySMHO with data sampled at a frequency \textbf{ten
  times} that of other methods. Additionally, DySMHO depends on the
  \texttt{pyomo} package, which interfaces with optimisation libraries
  like AMPL. Although the DySMHO utilised GAMS' version of
  \texttt{CONOPT}, we discovered that substituting \texttt{CONOPT} with
  \texttt{IPOPT} had minimal overall performance effect (see
  \cref{sec:conopt}). Owing to the constraints of the proprietary
  license regarding the permissible number of executions, we have
  selected \texttt{IPOPT} for the subsequent comparison.
\item   DAHSI: DAHSI exhibits insufficient performance to be relevant under
  the noise regime examined in this study; hence, a comprehensive
  comparison with DAHSI was not conducted. \Cref{sec:DAHSIvsODR}
  illustrates the comparison between DAHSI and ODR-BINDy.
\item   Ensemble-SINDy and modified-SINDy: The existing implementation uses a
  library that does not contain the constant term for the Lorenz63
  example. We have tested the inclusion of the constant term within the
  library; however, this action substantially diminishes the performance
  of the packages. Consequently, we adhere to the configuration provided
  by the packages and utilise solely the library without the constant
  term for Lorenz63.
\end{itemize}

We do note that, by using a slightly different dictionary for
Ensemble/modified-SINDy and a higher sampling frequency for DySMHO,
these packages confer certain advantages over others. Hence, the
following comparison should not be read as a fair benchmark
\textbf{between} existing packages. Instead, the comparison shows that,
despite favouring metrics, these existing packages do not outperform
ODR-BINDy in recovering the correct sparsity, i.e., the correct model,
in noisy and sparse datasets.

\subsection{Van Der Pol}\label{van-der-pol}

\begin{figure}

{\includegraphics[width=1.0\linewidth,height=\textheight,keepaspectratio]{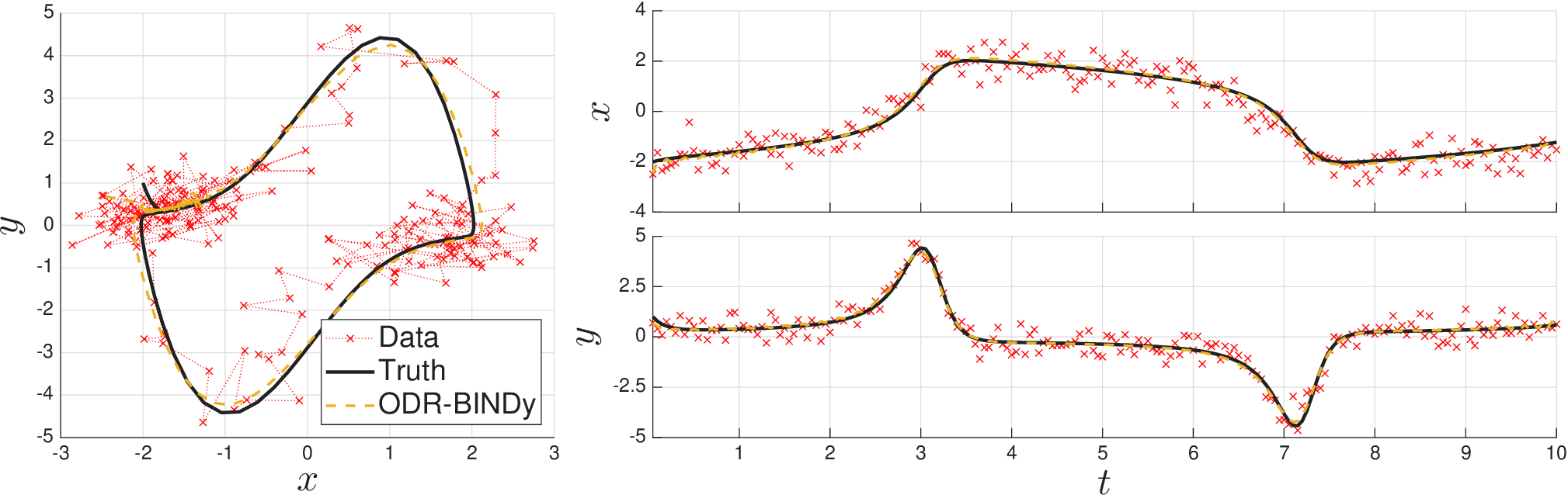}\label{fig:VanDerPol_Single}}

{\includegraphics[width=1.0\linewidth,height=\textheight,keepaspectratio]{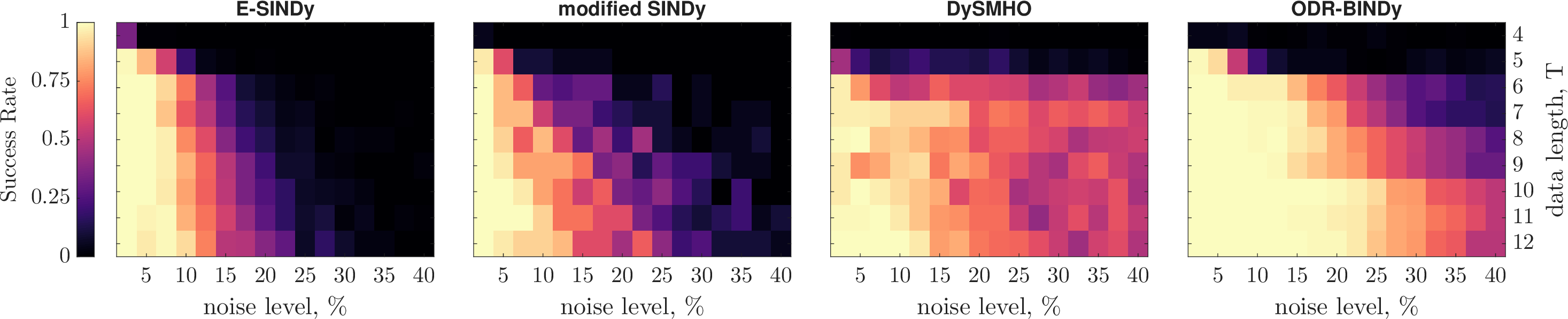}\label{fig:VanDerPol_Success}}

\caption[{(above) Example of recovering the Van Der Pol system and
denoising the data by ODR-BINDy. The data here contain \(30\%\) noise
and sampled at \(\Delta t=0.01\). (below) Success Rate of E-SINDy,
modified-SINDy, DySMHO and ODR-BINDy in recovering the Van Der Pol
system from data sampled at \(\Delta t=0.01\) (except for DySMHO, which
has \(\Delta t=0.001\)).}]{(above) Example of recovering the Van Der Pol
system and denoising the data by ODR-BINDy. The data here contain
\(30\%\) noise and sampled at \(\Delta t=0.01\). (below) Success Rate of
E-SINDy, modified-SINDy, DySMHO and ODR-BINDy in recovering the Van Der
Pol system from data sampled at \(\Delta t=0.01\) (except for DySMHO,
which has \(\Delta t=0.001\)).}

\label{fig:VanDerPol}

\end{figure}

\Cref{fig:VanDerPol_Single} shows Van Der Pol system \[
\dot{x}=y, \qquad \dot{y}=\mu(1-x^2)y-x, \qquad [x(0),y(0)]=[-2,1]
\] where \(\mu=0.5\). Sampled at \(\Delta t=0.01\) (\(\Delta t =0.001\)
for DySMHO) till the time \(T\) that specifies the data length, the data
is then contaminated with an i.i.d. Guassian Noise
\(\mathcal{N}(0,\sigma_x)\) before being fed into each package to
recover the governing equation and parameters based on a 3rd order
polynomial dictionary.

\Cref{fig:VanDerPol_Success} shows the performance of each package in
recovering the Van Der Pol system across data length, given by the time
\(T\), and the noise level. The noise level in the plot is defined as
the ratio \(\sigma_x/\sigma_{Signal}\), where \(\sigma_{Signal}\) is the
standard deviation of the flatten noiseless dataset. We see that
linear-based Ensemble-SINDy is not as noise-robust as other nonlinear
optimisation-based approaches. This is largely due to the nonlinear
approaches' ability to denoise data. Among the nonlinear methods,
ODR-BINDy is more noise robust and data efficient than other methods.
Meanwhile, DySMHO may seem slightly better performing than
modified-SINDy in a high noise regime, but note that DySMHO is using ten
times more data points to achieve comparable performance with
modified-SINDy and ODR-BINDy. This is likely due to the fact that it
uses the weak form collocation method to discretise the time domain. The
weak form tends to work well if the time window is highly sampled but
may also filter the critical signal in the high frequency if the
sampling frequency is not significantly higher than the signal \citep{Fung2025}.

\subsection{Lorenz 63}\label{lorenz-63}

\begin{figure}

{\includegraphics[width=1.0\linewidth,height=\textheight,keepaspectratio]{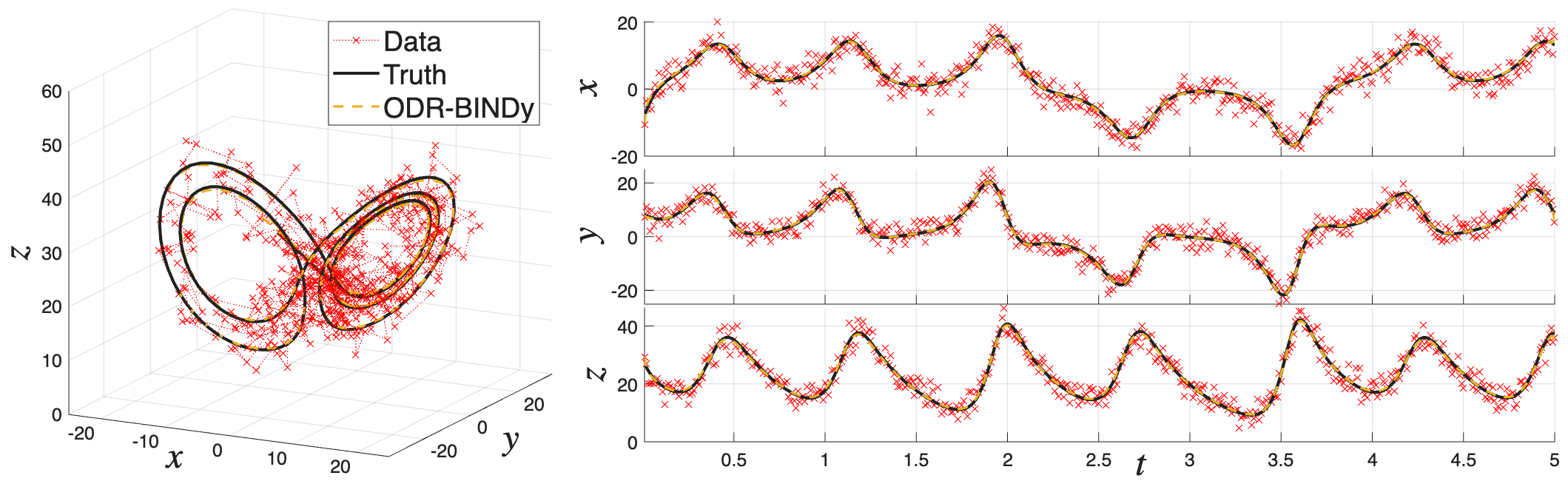}\label{fig:Lorenz_Single}}

{\includegraphics[width=1.0\linewidth,height=\textheight,keepaspectratio]{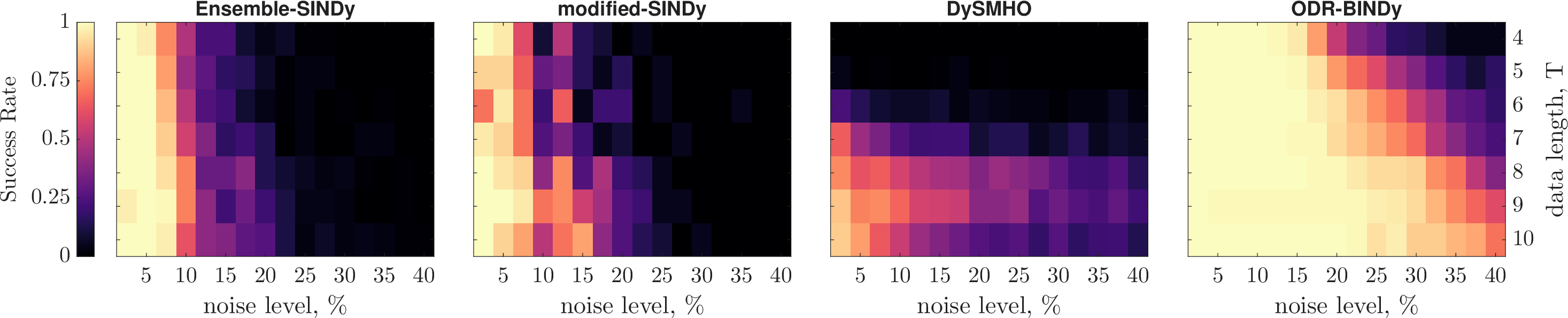}\label{fig:Lorenz_Success}}

\caption[{(above) Example of recovering the Lorenz63 system and
denoising the data by ODR-BINDy. The data here contain \(20\%\) noise
and sampled at \(\Delta t=0.01\). (below) Success Rate of E-SINDy,
modified-SINDy, DySMHO and ODR-BINDy in recovering the Lorenz63 system
from data sampled at \(\Delta t=0.01\) (except for DySMHO, which has
\(\Delta t=0.001\)).}]{(above) Example of recovering the Lorenz63 system
and denoising the data by ODR-BINDy. The data here contain \(20\%\)
noise and sampled at \(\Delta t=0.01\).  (below) Success Rate of E-SINDy,
modified-SINDy, DySMHO and ODR-BINDy in recovering the Lorenz63 system
from data sampled at \(\Delta t=0.01\) (except for DySMHO, which has
\(\Delta t=0.001\)).}

\label{fig:Lorenz}

\end{figure}

\Cref{fig:Lorenz} shows the typical Lorenz63 system \[
\dot{x}=\sigma(x-y),\qquad \dot{y}=x(\rho-z)-y,\qquad \dot{z}=xy-\beta z \qquad [x(0),y(0),z(0)]=[-8,8,27]
\] with \(\sigma=10,\rho=28,\beta=8/3\), and the performance of each
package in recovering the system from data sampled as \(\Delta t= 0.01\)
(\(\Delta t= 0.001\) for DySMHO), contaminated by Gaussian noise. A 2nd
order polynomial library is used, but as described above, an exception
was made for Ensemble-SINDy and modified-SINDy to remove the constant
term from the library to improve their convergence. Again, we see that
linear-based Ensemble-SINDy is not as noise-robust as the nonlinear
optimisation-based approaches. However, modified-SINDy and DySMHO are
also not performing as well as in the Van Der Pol system. This can be
attributed to the fact that Lorenz63 is chaotic, and therefore
modified-SINDy and DySMHO cannot fully make use of the whole dataset in
recovering the equation. Instead, DySMHO limits the scope of the data by
learning from a limited horizon, while modified-SINDy penalises losses
from data too far behind or into the future from the current time. Both
of these strategies limit the information one can gain from the data and
therefore limit noise robustness and data efficiency. In contrast,
ODR-BINDy can assimilate the entire time series to learn the model and
can therefore maximise the information one can extract from the data
regarding the parameter values and model selection.

\subsection{R{\"o}ssler}\label{sec:Rossler}

\begin{figure}

{\includegraphics[width=1.0\linewidth,height=\textheight,keepaspectratio]{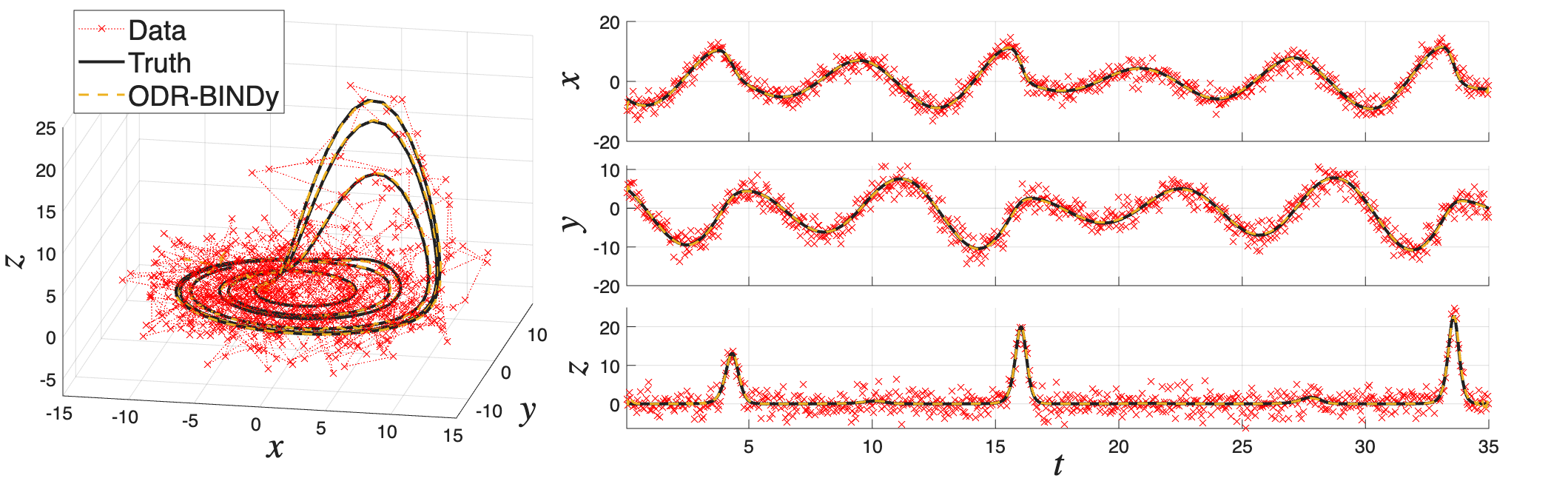}\label{fig:Rossler_Single}}

{\includegraphics[width=1.0\linewidth,height=\textheight,keepaspectratio]{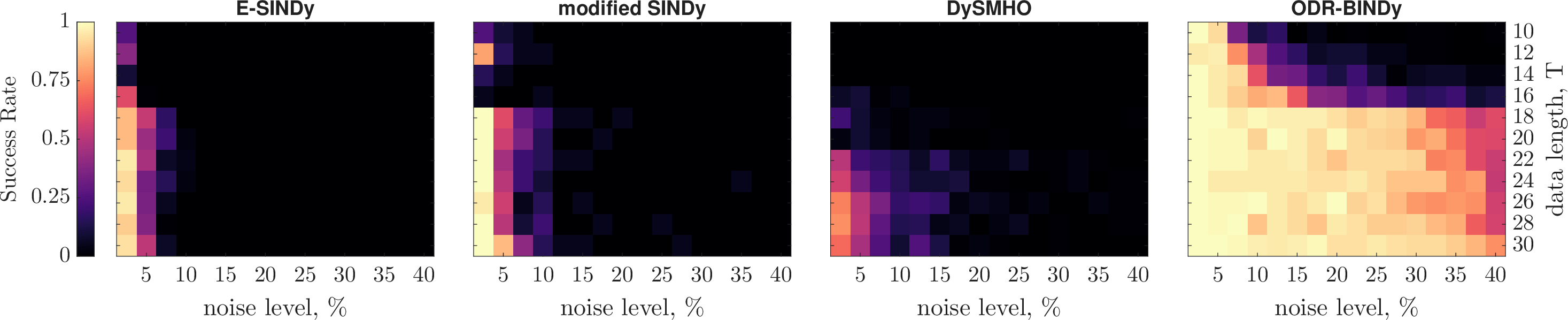}\label{fig:Rossler_Success}}

\caption[{(above) Example of recovering the R{\"o}ssler system and denoising
the data by ODR-BINDy. The data here contain \(40\%\) noise and sampled
at \(\Delta t=0.05\). (below) Success Rate of E-SINDy, modified-SINDy,
DySMHO and ODR-BINDy in recovering the R{\"o}ssler system from data sampled
at \(\Delta t=0.05\) (except for DySMHO, which has
\(\Delta t=0.005\)).}]{(above) Example of recovering the R{\"o}ssler system
and denoising the data by ODR-BINDy. The data here contain \(40\%\)
noise and sampled at \(\Delta t=0.05\). (below) Success Rate of E-SINDy,
modified-SINDy, DySMHO and ODR-BINDy in recovering the R{\"o}ssler system
from data sampled at \(\Delta t=0.05\) (except for DySMHO, which has
\(\Delta t=0.005\)).}

\label{fig:Rossler}

\end{figure}

\Cref{fig:Rossler_Single} shows the same comparison in the R{\"o}ssler
system \[
\dot{x}=-y-z,\qquad \dot{y}=x+ay,\qquad \dot{z}=b+z(x-c),\qquad [x(0),y(0),z(0)]=[-6,5,0]
\] where \(a=0.2,b=0.2,c=5.7\). The data is sampled at
\(\Delta t=0.05\), and contaminated by Gaussian noise scaled by the
noise level. In this case, all package are attempting to learn the model
from a 2nd order polynomial library with constant term. Each package's
performance is shown in \cref{fig:Rossler_Success} . Similar to the
results before, ODR-BINDy is more noise-robust and data-efficient than
other methods.

\section{Limitations}\label{limitations}

\begin{figure}
\centering
\includegraphics[width=1.0\linewidth,height=\textheight,keepaspectratio]{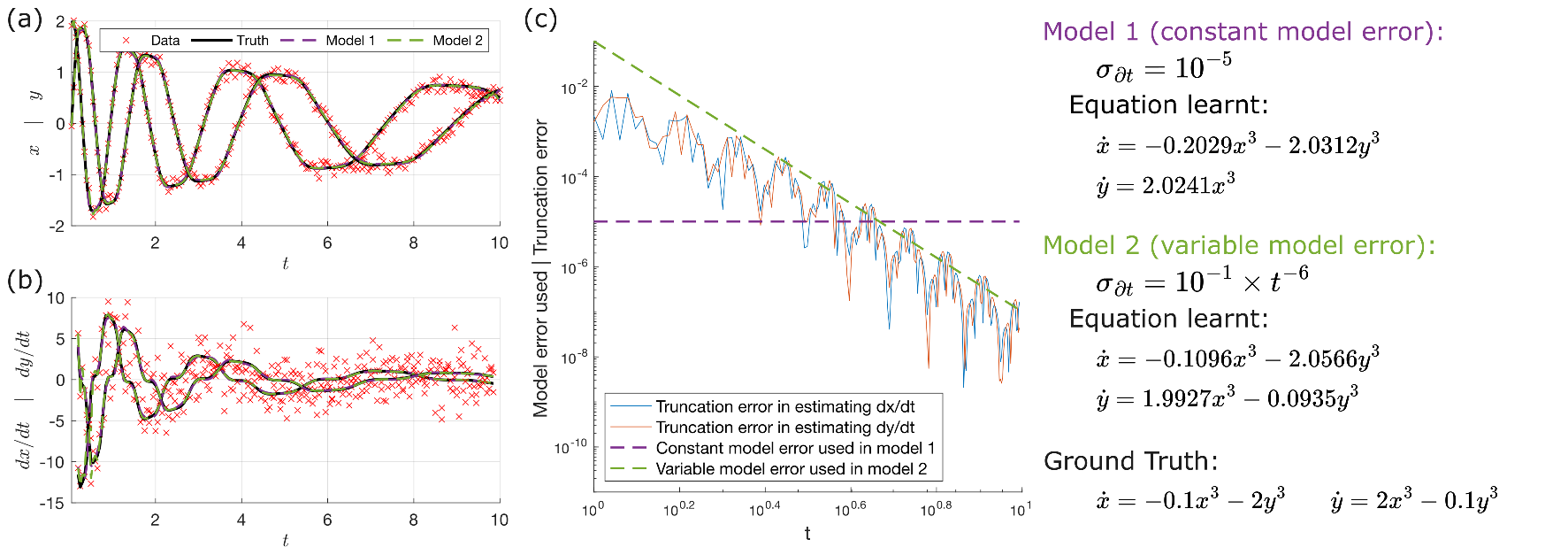}
\caption{Attempts to learn the equation of the nonlinear oscillator
using ODR-BINDy. Here, \(\Delta t = 0.05\) and \(\sigma_x=0.084\)
(\(10\%\) noise). Since both \((a)\) the states and \((b)\) the time
derivatives are decaying, \((c)\) the truncation error is also decaying
in time. Here, we compare two models, with different hyperparameter
settings for \(\sigma_{\partial t}\) to demonstrate the potential issue
in learning from decaying signals. Model 1 (purple) represents the
typical hyperparameter setting for \(\sigma_{\partial t}\), which is not
consistent with the real truncation error. Model 2 (green) assumes that
the trend of the truncation error is somehow known \emph{a priori}.
Although both models seemingly able to follow the signal in \((a-b)\),
the equation learnt by model 1 is oversimplified, likely owing to the
fact that the noise is overestimated later in time. Only by modifying
the noise to follow the real truncation error in model 2 can the correct
model be inferred, but in practice the truncation error is likely not
known \emph{a priori}.}\label{fig:NonlinOsc}
\end{figure}

Although ODR-BINDy generally shows better performance than other
existing packages in recovering the correct model, especially in data
contaminated by large measurement noise, there remain some limitations
to ODR-BINDy:

\begin{itemize}
\item  The performance of ODR-BINDy is not as good in decaying signals, such
  as damped cubic oscillators. This is due to the fact that the
  discretisation error is currently modelled as a time-invariant
  stochastic noise, while a decaying signal will lead to a decaying
  discretisation error that temporally correlates with the signal
  itself. Such time-varying and signal-correlating errors could lead to
  bias in the model and thereby prevent the Bayesian algorithm from
  selecting the correct model. \Cref{fig:NonlinOsc} shows an example of
  ODR-BINDy failing to recover the cubic oscillator equation from its
  decaying signal.
\item  Currently, ODR-BINDy is not designed to directly provide forecasts of
  the time-series data. Works are ongoing in demonstrating ODR-BINDy's
  capability for direct forecasting, but there are some technical
  challenges in ensuring the statistical guarantee of the forecast as
  one switches from the finite difference time discretisation to
  conventional ODE solvers for the forecasting. There is also the subtle
  challenge of how one should determine the distribution of the initial
  condition for the forecast from the given historical data. With that
  said, one can always apply classical data assimilation techniques
  \citep{Asch2016} on the model/sparsity learnt from ODR-BINDy to
  perform forecasting. The procedure of this two-step strategy will be
  similar to many classical works on data assimilation and forecasting,
  so we shall not repeat them here.
\end{itemize}

\section{Conclusion and future work}\label{conclusion-and-future-work}

This study demonstrates the efficacy of ODR-BINDy in identifying
governing equations of a dynamical system from noisy temporal state
observations. It has not only exhibited superior noise robustness and
data efficiency compared to all current SINDy variants in accurately
recovering the correct model, but its speed also surpasses that of all
existing SINDy variants that utilise nonlinear optimisation to address
the error-in-variable problem. A portion of the performance improvement
can presumably be ascribed to the Bayesian framework, which offers a
robust basis for model selection and the enhancement of sparsity.
However, the chief breakthrough that gives rise to the algorithm's speed
and capacity to assimilate long time series even in chaotic systems is
due to stochastic relaxation. Stochastic relaxation not only mitigated
the sensitivity issue associated with conventional parameter estimation
in chaotic systems but also underscored the significance of truncation
errors in data assimilation. Furthermore, it elucidated the origins of
the scaling factors for each component of the total loss function,
providing both intuitive guidance for tuning and theoretical rationale
for the associated hyperparameters.

Despite the algorithm's demonstrated efficiency, partly due to the
well-optimised \texttt{lsqnonlin} algorithm and the implementation of
sparse and analytical gradients, there exists considerable potential for
enhancement, particularly regarding its computing speed. The existing
greedy approach exhibits a complexity of \(\mathcal{O}(NDb^2 M^2)\),
where \(M\) represents the number of candidate functions in the library
and increases exponentially with \(D\). The quadratic scaling is a result 
 of the current greedy algorithm, which evaluates the effect of the
removal of each term present in the model. The computational complexity
can be reduced to \(\mathcal{O}(NDb^2 M)\) if we incorporate analytical
estimation of the potential change in evidence for the removal testing
process, akin to the strategy used by the \texttt{SparseBayes} method
\citep{Tipping2003}. This improvement will be essential for
high-dimensional applications, such as the recovery of dynamics in
reduced-order models for fluid problems \citep{Kaptanoglu2021}.
Alternatively, employing gradient descent methods such as Adam in place
of the existing multi-start trust-region approach may enhance
performance if the initial guess does not give satisfactory results.

Besides potential improvement in computational speed, there are many
aspects where the method can be further improved. For instance, although
we have solely illustrated the method's efficacy utilising an equispaced
finite difference discretisation, numerous collocation methods,
including the weak formulation, exist to discretise the equation (both
equi-spatially and non-equispatially) that can also be represented as
the matrices \(L_{\partial t}\) and \(L_{I}\). Consequently, one can
readily interchange these matrices with those associated with the
preferred collocation or discretisation technique. Alternatively, one
may pursue high-order ODE solvers to impose the equation constraint, as
exemplified by modified-SINDy \citep{Kaheman2022}, rather than use
the discretize-then-optimize methodology. Nevertheless, additional
effort is necessary to integrate stochastic relaxation into this
methodology.

Moreover, our formulation in \cref{eq:loss_soft} can be readily adapted
to accommodate non-Gaussian noise. In particular, further research is
required to examine the actual noise distribution resulting from
truncation errors caused by discretisation. Although relaxing the noise
model to a non-Gaussian distribution will lead to a non-Gaussian
posterior, in the usual case where the number of data points far exceeds
the number of parameters, the posterior is most likely unimodal and
sub-Gaussian. Hence, the Gauss-Newton-based Laplace approximation of the
evidence is likely a good approximation for most cases.

Lastly, the nonlinear optimisation of our current algorithm also allows
for a more extensible dictionary. Indeed, part of DySMHO
\citep{Lejarza2022} and DAHSI \citep{Ribera2022} have already
demonstrated learning a parameterised library, and extending the current
algorithm to include an implicit term is also possible.
\citet{Ribera2022} have also demonstrated the possibility of
recovering the full equations with some of the state variable
measurements hidden from the algorithm. It would be interesting to
investigate if ODR-BINDy is capable of similar problems, although it is
unclear if the algorithm can perform just as well without the
observational data as part of the initial condition.

\backmatter

\bmhead{Acknowledgements} We would like to to thank U. Fasel, A. Novoa, L. Magri, B. Moseley and M. P. Juniper for their feedback on the manuscript, and J. N. Kutz for the brief but fruitful discussion.

\bmhead{Code availability} The code for ODR-BINDy can be found on \href{https://github.com/llfung/ODR-BINDy}{github.com/llfung/ODR-BINDy}. The code to generate the comparisons with other methods can be found on the following links: \href{https://github.com/llfung/ODR-BINDy-EnsembleSINDy-Benchmark}{E-SINDy}, \href{https://github.com/llfung/ODR-BINDy-modified-SINDy-Benchmark}{modified-SINDy} and \href{https://github.com/llfung/ODR-BINDy-DySMHO-Benchmark}{DySMHO}.

\bmhead{Data availability} The data on the success rate and model error coefficients can be found on \href{https://doi.org/10.5281/zenodo.16614612}{Zenodo}.

\begin{appendices}
\section{Appendix}\label{sec:Appendix}

\subsection{Averaged model parameter errors}\label{sec:param_error}

\begin{figure}
\centering
\includegraphics[width=1.0\linewidth,height=\textheight,keepaspectratio]{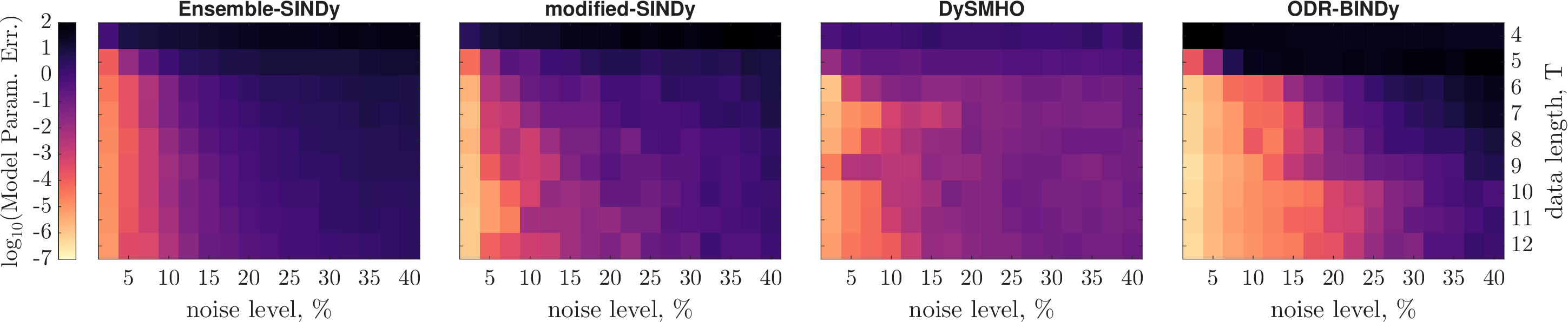}
\caption{Van Der Pol: Comparison of model parameter error from the
parameter values learnt by E-SINDy, modified-SINDy, DySMHO and ODR-BINDy
at increasing noise-to-signal ratio and decreasing data length.
Hyperparmeter and settings are the same as that of \cref{fig:VanDerPol}
for each package.}\label{fig:VanDerPol_ParamError}
\end{figure}

\begin{figure}
\centering
\includegraphics[width=1.0\linewidth,height=\textheight,keepaspectratio]{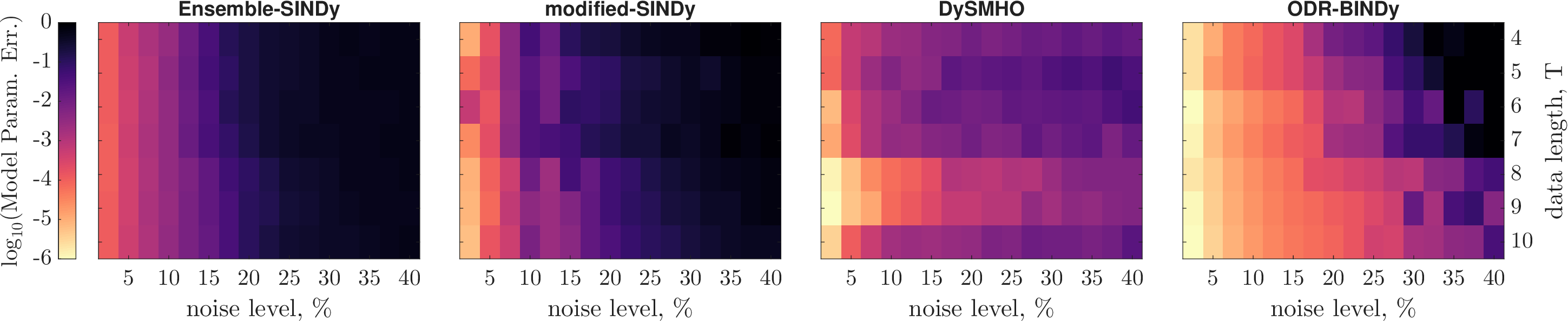}
\caption{Lorenz63: Comparison of model parameter error from the
parameter values learnt by E-SINDy, modified-SINDy, DySMHO and ODR-BINDy
at increasing noise-to-signal ratio and decreasing data length.
Hyperparmeter and settings are the same as that of \cref{fig:Lorenz} for
each package.}\label{fig:Lorenz_ParamError}
\end{figure}

\begin{figure}
\centering
\includegraphics[width=1.0\linewidth,height=\textheight,keepaspectratio]{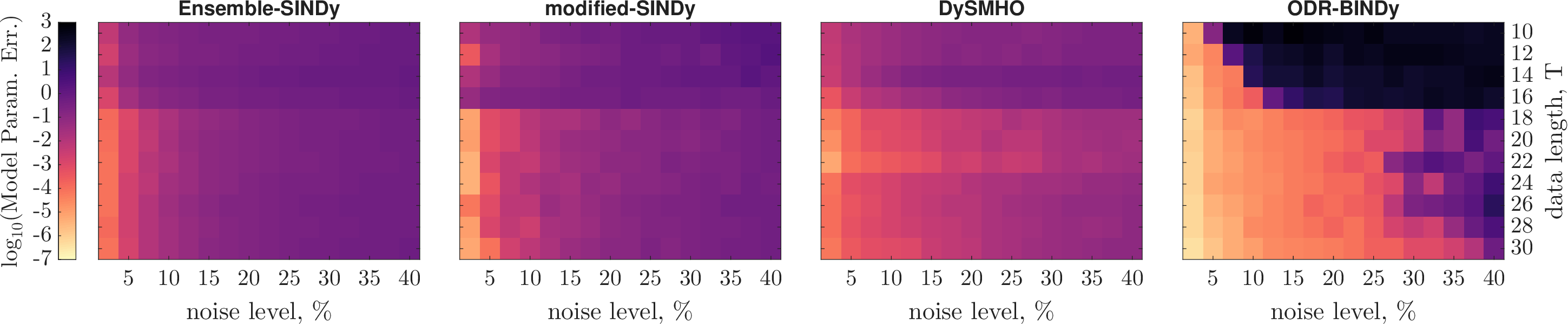}
\caption{R{\"o}ssler: Comparison of model parameter error from the parameter
values learnt by E-SINDy, modified-SINDy, DySMHO and ODR-BINDy at
increasing noise-to-signal ratio and decreasing data length.
Hyperparmeter and settings are the same as that of \cref{fig:Rossler}
for each package.}\label{fig:Rossler_ParamError}
\end{figure}

\Cref{fig:VanDerPol_ParamError}, \cref{fig:Lorenz_ParamError} and
\cref{fig:Rossler_ParamError} show the averaged model parameter error
from the learning result of Ensemble-SINDy, modified-SINDy, DAHSI and
ODR-BINDy. The model parameter error of each run is defined by the
normalised difference between the parameter value learnt against the
ground truth parameter value \(\boldsymbol\Xi_{true}\), i.e. \[
\frac{||\boldsymbol\Xi-\boldsymbol\Xi_{true}||_F}{||\boldsymbol\Xi_{true}||_F}
\] and the plots show the averaged value over many runs. (Note that
despite the Frobenius norm used here being different from the 2-norm
used in (17) of \citep{Kaheman2022}, the vector \(\Xi\) in
\citep{Kaheman2022} was assumed to be flatten. Hence, the parameter
error defined here is the same as (17) in \citep{Kaheman2022}. ) The
hyperparameter setting are the same as the one used in previous
sections, and each plot corrsponds to the respective success rate plot
in \cref{fig:VanDerPol}, \cref{fig:Lorenz} and \cref{fig:Rossler}.

\subsection{Formula for the Laplace Approximation in
ODR}\label{sec:EvidenceFormula}

In Orthogonal Distance Regression (ODR), we seek to minimise the loss
function as specified in \cref{eq:loss_soft} by varying \(\mathbf{X}\)
and \(\boldsymbol{\Xi}\). To facilitate easier notation in the following
derivation, we define
\(\eta_{ij} = \sum_{k,n}[L_{\partial t}]_{ik} {X}_{kj} -  [L_{I}]_{ik} \Theta_{kn}\Xi_{nj}\)
and \(\zeta_{ij}=X_{ij}-\hat{X}_{ij}\). Therefore, \cref{eq:loss_soft}
can be rewritten in index form as \[
\mathcal{L}(\boldsymbol\Xi,\mathbf{X}) =\frac{1}{2}\left[\sum_{id}\frac{\zeta_{id}^2}{\sigma^2_{x}}+\sum_{id}\frac{\eta_{id}^2}{\sigma^2_{\partial t}}+\sum_{mn}\frac{\Xi_{mn}^2}{\sigma^2_{p}}\right]
\]

In particular, by minimising the above against \(\mathbf{X}\), we are
also effectively minimising
\({\zeta_{ij}^2}/{\sigma^2_{x}}+{\eta_{ij}^2}/{\sigma^2_{\partial t}}\),
the ``rescaled'' distance between the data points \(\hat{\mathbf{X}}\)
and the respective point \(\mathbf{X}\) on the regressed curve, the
rescaling of which is defined by \(\sigma_x\) and
\(\sigma_{\partial t}\) on the data error space (spanned by
\(\boldsymbol\zeta\)) and model error space (spanned by
\(\boldsymbol\eta\)) respectively. The resulting optimal points
\(\mathbf{X}=\mathbf{X}^*(\boldsymbol{\Xi})\) is the ``rescaled closest
points'' on the curve to the respective data point \(\hat{\mathbf{X}}\)
that not only minimises \(\mathcal{L}(\boldsymbol\Xi,\mathbf{X})\) for a
given curve specified by \(\boldsymbol{\Xi}\), they can also be
interpreted as the corresponding estimated denoised states.

To help with the notation and the following derivation, we shall define
the following derivatives: \[
\eta_{id}=\sum_{k,n}[L_{\partial t}]_{ik} {X}_{kd} -  [L_{I}]_{ik} {\Theta}_{kn}\Xi_{nd}
\] \[
\frac{\partial \eta_{id}}{\partial X_{je}}=\sum_{k,n}[L_{\partial t}]_{ij}\delta_{de}-  [L_{I}]_{ik} \frac{\partial \Theta_{kn}}{\partial X_{je}}\Xi_{nd}
\] \[
\frac{\partial^2 \eta_{id}}{\partial X_{je}\partial X_{lf}} = \sum_{k,n}-  [L_{I}]_{ik} \frac{\partial \Theta_{kn}}{\partial X_{je}\partial X_{lf}}\Xi_{nd}
\] \[
\frac{\partial \eta_{id}}{\partial \Xi_{ne}}= \sum_{k}- [L_{I}]_{ik} {\Theta}_{kn}\delta_{de}=\frac{\partial \eta_{id}}{\partial \Xi_{nd}}
\] \[
\frac{\partial^2 \eta_{id}}{\partial \Xi_{ne}\partial X_{jf}}= \sum_{k}- [L_{I}]_{ik} \frac{\partial \Theta_{kn}}{\partial X_{jf}} \delta_{de}=\frac{\partial^2 \eta_{id}}{\partial \Xi_{nd}\partial X_{jf}}.
\]

Now, since \(\mathbf{X}^*\) is the optimal point, one can show that at
\(\mathbf{X}=\mathbf{X}^*(\boldsymbol{\Xi})\) and any
\(\boldsymbol{\Xi}\),
\begin{equation}\protect\phantomsection\label{eq:orthogonal}{
\left. \frac{\partial\mathcal{L}}{\partial {X}_{je}} \right|_{\mathbf{X}=\mathbf{X}^*,\boldsymbol{\Xi}} = \frac{\zeta_{je}}{\sigma_x^2}+ \frac{\eta_{id}}{\sigma_{\partial t}^2} \frac{\partial \eta_{id}}{\partial X_{je}}=\mathbf{0}
}\end{equation}

This is the orthogonal constrain that underly ODR \citep[see
fig.~1-2]{Boggs1987}, which has a geometrical interpretation as follow:
at the optimal point \(\mathbf{X}=\mathbf{X}^*\) that minimises the
rescaled distance, the rescaled distance vector
\([\boldsymbol\zeta/\sigma_x;\boldsymbol\eta/\sigma_{\partial t}]\) must
be orthogonal to the rescaled local slope of the curve at
\(\mathbf{X}^*\), which is
\([\mathbf{1}/\sigma_x;(\partial\boldsymbol\eta/\partial \mathbf{X})/\sigma_{\partial t}]\).
While in modern implementation of ODR, we no longer use the above
constrain to compute \(\mathbf{X}^*\), here, we shall assume that when
the nonlinear regression converges to the optimal point, the solution
must also satisfy \cref{eq:orthogonal}. We shall use this fact as the
basis to derive the analytical expression of the Hessian needed to
compute the Bayesian evidence.

Since the optimisation to minimise \(\mathcal{L}\) will inevitably gives
a solution where \(\mathbf{X}=\mathbf{X}^*\), from now on, we will
assume the optimal \(\mathbf{X}\) is computed. Here, we further simplify
the expression of all terms that is computed at
\(\mathbf{X}=\mathbf{X}^*(\boldsymbol\Xi)\) with the \(^*\) notation,
e.g.~\(\eta_{id}^*=\eta_{id}|_{X=X^*}\).

Differentiating \cref{eq:orthogonal} by \(\boldsymbol{\Xi}\) gives 
\begin{align}
      & \left(\frac{1}{\sigma_x^2}\delta_{de}\delta_{ij}+ \frac{1}{\sigma_{\partial t}^2}(\frac{\partial \eta_{kf}^*}{\partial X_{je}} \frac{\partial \eta_{kf}^*}{\partial X_{id}} +\eta_{kf}^* \frac{\partial^2 \eta_{kf}^*}{\partial X_{je}\partial X_{id}} ) \right)\frac{dX_{je}^*}{d\Xi_{ne}} \nonumber \\
      = & \frac{-1}{\sigma_{\partial t}^2}\left( \frac{\partial \eta_{kf}^*}{\partial \Xi_{ne}} \frac{\partial \eta_{kf}^*}{\partial X_{id}}+\eta_{kf}^*\frac{\partial^2 \eta_{kf}^*}{\partial \Xi_{ne}\partial X_{id}}\right),
\end{align}
which we can solve for \({dX_{je}^*}/{d\Xi_{ne}}\) numerically upon
finding \(\mathbf{X}^*\) from the nonlinear optimisation. With
\({dX_{je}^*}/{d\Xi_{ne}}\) and the above derivatives computed, we are
now ready to calculate the Bayesian evidence.

\subsubsection{Computing the evidence by Gauss-Newton Laplace
approximation}\label{computing-the-evidence-by-gauss-newton-laplace-approximation}

Now, to compute the Bayesian evidence by Laplace approximation, we
require the Hessian 
\begin{align}
\frac{d^2\mathcal{L}}{d{\Xi_{nd}}d{\Xi_{me}}} & = \frac{1}{\sigma_{\partial t}^2}\left[\eta^*_{kf}(\frac{\partial^2 \eta_{kf}^*}{\partial \Xi_{nd}\partial X_{ig}}\frac{dX_{ig}^*}{d\Xi_{me}}+\frac{\partial^2 \eta_{kf}^*}{\partial \Xi_{me}\partial X_{ig}}\frac{dX_{ig}^*}{d\Xi_{nd}}+\frac{\partial\eta^*_{kf}}{\partial X_{ig}}\frac{d^2X^*_{ig}}{d\Xi_{nd} d\Xi_{me}}) \right. \\
 & + \left. (\frac{\partial\eta^*_{kf}}{\partial X_{ig}}\frac{dX_{ig}^*}{d\Xi_{nd}}+\frac{\partial \eta^*_{kf}}{\partial \Xi_{nd}})(\frac{\partial\eta^*_{kf}}{\partial X_{ig}}\frac{dX_{ig}^*}{d\Xi_{me}}+\frac{\partial \eta^*_{kf}}{\partial \Xi_{me}})\right] \\
& + \frac{1}{\sigma_x^2}\left[ (X_{ig}-\hat{X}_{ig})\frac{d^2 X^*_{ig}}{d\Xi_{nd} d\Xi_{me}}+\frac{dX^*_{ig}}{d\Xi_{nd}}\frac{dX^*_{ig}}{d\Xi_{me}}\right]+\frac{1}{\sigma_p^2}\delta_{nm} \delta_{de}
\end{align}
which involves expensive computation of
\(d^2 \mathbf{X} /d\boldsymbol{\Xi} d\boldsymbol{\Xi}\). The
Gauss-Newton approximation allows us to approximate the above Hessian
without computing
\(d^2 \mathbf{X} /d\boldsymbol{\Xi} d\boldsymbol{\Xi}\), by effectively
ignoring the second order derivatives, giving
\begin{align}
\frac{d^2\mathcal{L}}{d \Xi_{nd} d\Xi_{me}} &\approx  
\frac{1}{\sigma_{\partial t}^2} \left[ (\frac{\partial\eta^*_{kf}}{\partial X_{ig}}\frac{dX_{ig}^*}{d\Xi_{nd}}+\frac{\partial \eta^*_{kf}}{\partial \Xi_{nd}})(\frac{\partial\eta^*_{kf}}{\partial X_{ig}}\frac{dX_{ig}^*}{d\Xi_{me}}+\frac{\partial \eta^*_{kf}}{\partial \Xi_{me}}) \right] \nonumber \\
& +  \frac{1}{\sigma_x^2}\left[ \frac{dX^*_{ig}}{d\Xi_{nd}}\frac{dX^*_{ig}}{d\Xi_{me}}\right]+\frac{1}{\sigma_p^2}\delta_{nm} \delta_{de} 
\end{align}
Plugging in the derivatives computed in the last section for each
term above, we can approximate the Hessian above and compute the log
Bayesian evidence by \[
\mbox{Log-Evidence} = - \left( \mathcal{L}(\Xi^*,\mathbf{X}^*)+\frac{N_\Xi}{2}(\log{(2\pi)}+2\log{\sigma_p})+\frac{1}{2}\log \left| \frac{1}{2\pi}\frac{d^2\mathcal{L}}{d\boldsymbol{\Xi}d\boldsymbol{\Xi}} \right| \right)
\] in which \(N_\Xi\) is the number of active terms in the model.
Lastly, the Greedy algorithm is used to find the model that maximises
the log-evidence above.

\subsection{Comparison with DAHSI}\label{sec:DAHSIvsODR}

\begin{figure}
\centering
\includegraphics[width=0.9\linewidth,height=\textheight,keepaspectratio]{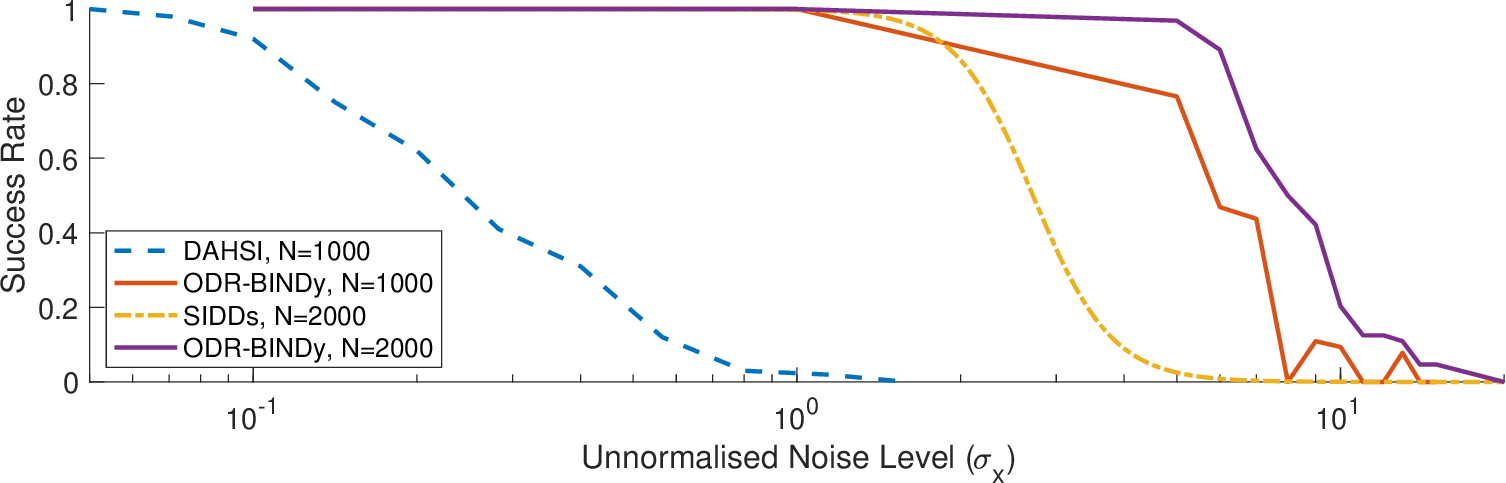}
\caption{Success rate of DAHSI, SIDDs and ODR-BINDy in recovering the
Lorenz63 model from 1000/2000 data points, sampled at \(\Delta t=0.01\).
Here, the hyperparameters used in DAHSI are \(\alpha=1.1\),
\(\beta_{\max}=256\), \(\lambda=0.05:0.05:1.01\), whereas the data of
SIDDs is a rough estimate from fig 5.4 of \citep{Hokanson2023}. Data
of ODR-BINDy are generated using the same settings as before. Note that
the metric for ``success'' for DAHSI is more generous than the metric
for ODR-BINDy. In all cases, ODR-BINDy is shown to be more data
efficient and noise robust in recovering Lorenz63 from data than DAHSI
and SIDDs.}\label{fig:DAHSI}
\end{figure}

The main DAHSI package generates many models from sweeping the
thresholding (\(\lambda\)) and Variational Annealing (\(\alpha^\beta\))
hyperparameters, from which a model is selected semi-automatically by
information criterion or Pareto front. Given that certain manual inputs
are anticipated in the model selection procedure, a direct comparison
with ODR-BINDy on its capacity for automatic model recovery may not be
the most appropriate. Nevertheless, as a highly generous metric, one can
calculate the rate of identifying the true model in \textbf{at least one
model} produced by the package, which is equivalent to assuming that the
subsequent model selection process via AIC/BIC or Pareto front will
inevitably choose the correct model from the parameter sweep. Despite
the favourable metric, \cref{fig:DAHSI} shows that DAHSI cannot recover the original model most of the time in the noise regime examined by this work. Instead, DAHSI works
only in a very low noise regime.

\subsection{Comparison with SIDDs}\label{comparison-with-sidds}

Although we were unable to reproduce the results of SIDDs due to absent
dependencies, we made every effort to compare our findings with the
compiled plots shown in the original SIDDs publication. Upon
approximating the result of Fig. 5.4 in \citep{Hokanson2023} in
\cref{fig:DAHSI}, which illustrates the success rate in retrieving the
same Lorenz63 equations using 2000 data points sampled at
\(\Delta t = 0.01\), we assert that ODR-BINDy demonstrates superior
performance compared to SIDDs. This is likely attributable to the
stochastic relaxation and Bayesian approach utilised in this study.

\subsection{Gauss-Newton Approximation}\label{sec:GaussNewton}

\begin{figure}
\centering
\includegraphics[width=0.9\linewidth,height=\textheight,keepaspectratio]{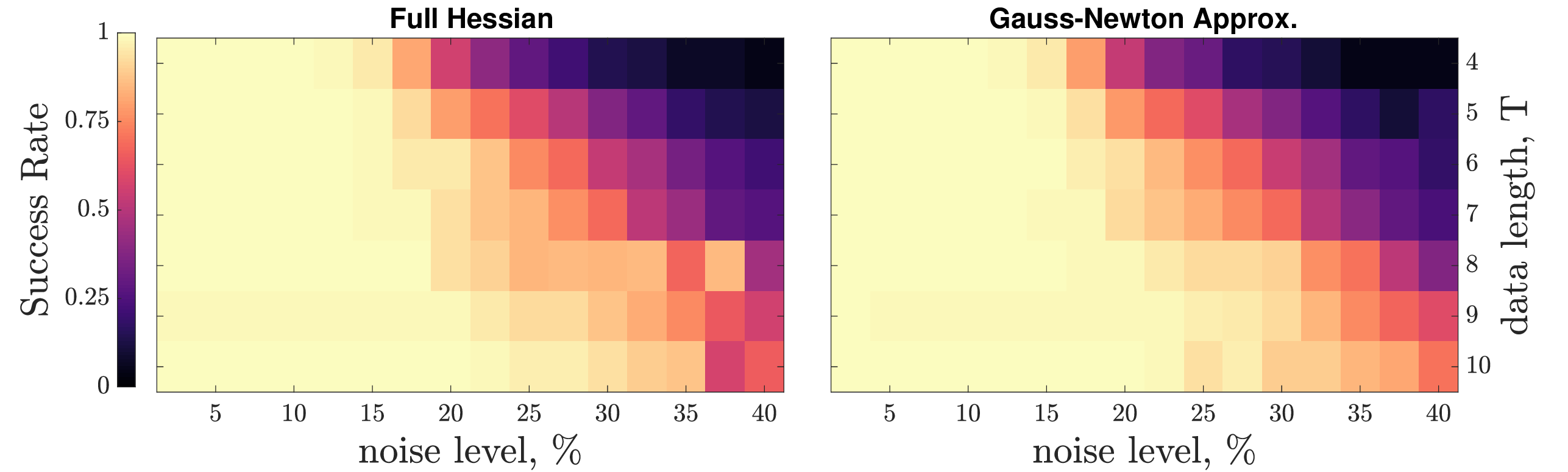}
\caption{Comparison of success rate of ODR-BINDy  in recovering the Lorenz63 model sampled at \(\Delta t=0.01\) using the full computation of the Hessian (left) and the Gauss-Newton approxmation (right) for the Laplace Approximation of the Bayesian evidence, the metric used to score different models. The use of Gauss-Newton have little effect on the overall performance. The hyperparameters used here is the same as \cref{fig:Lorenz}. } \label{fig:GaussNewton}
\end{figure}

\Cref{fig:GaussNewton} shows the performance comparison between employing the
full evidence approximation (with Hessian) that the Gauss-Newton
approximation of the evidence (without Hessian). The use of Gauss-Newton
approximation seems to have little impact on the overall success rate of
recovering the true model, but the performance gain of avoiding the full
Hessian computation is substantial.

\subsection{\texorpdfstring{DySMHO \texttt{CONOPT} vs
\texttt{IPOPT}}{DySMHO CONOPT vs IPOPT}}\label{sec:conopt}

\begin{figure}
\centering
\includegraphics[width=0.90\linewidth,height=\textheight,keepaspectratio]{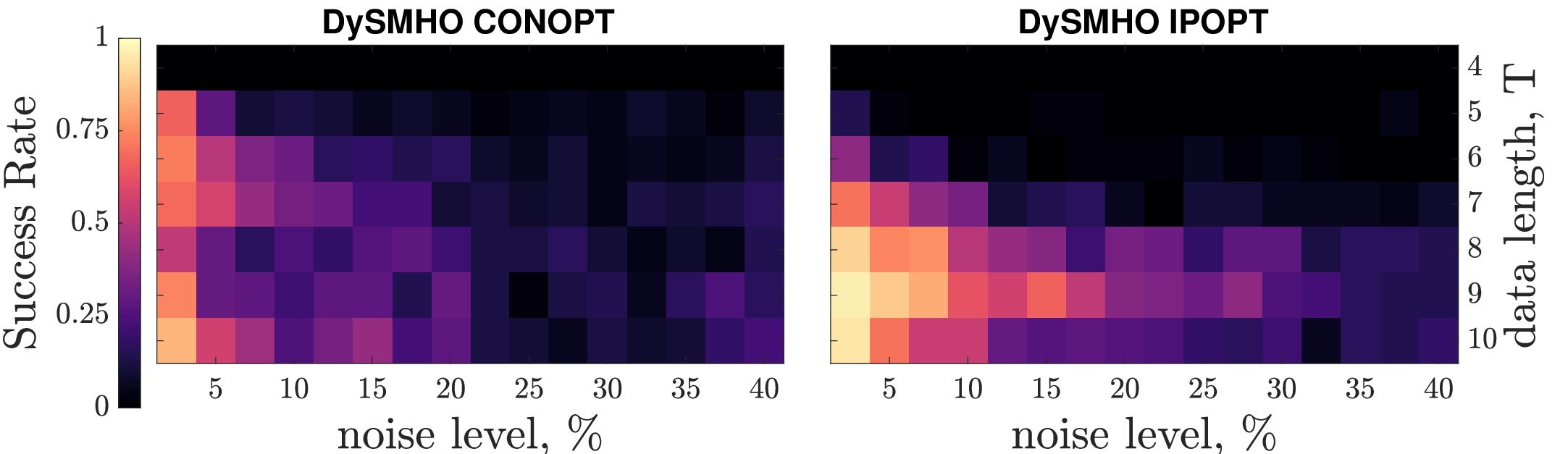}
\caption{Comparison of success rate of DySMHO in recovering the Lorenz63
model sampled at \(\Delta t=0.001\) using \texttt{CONOPT} (left) and
\texttt{IPOPT} (right). Here, the hyperparameters used are: time window
= 2 time unit, time step between window = 100, \texttt{nfe}=50,
\texttt{ncp}=5, thresholding frequency =10 and thresholding tolerance=1.
Notice that \texttt{IPOPT} is slightly better performing than
\texttt{CONOPT}.}\label{fig:conopt_ipopt}
\end{figure}

Due to licensing issue, we have only had a brief period of time where we
have access to GAM's \texttt{CONOPT} that was used in the original work
\citep{Lejarza2022} . To work around the issue, we have replaced
\texttt{CONOPT} with the open source \texttt{IPOPT} with \texttt{HSL}.
To make sure the swap doesn't induce a significant performance change,
we have compared the two methods. The comparison is shown in
\cref{fig:conopt_ipopt}, which shows that the replacement of
\texttt{CONOPT} wtih \texttt{IPOPT} actually slightly improve overall
performance.

\subsection{Hyperparameters of each
algorithms}\label{hyperparameters-of-each-algorithms}
\Cref{tbl:hyperparameters} listed all the hyper-parameters used by each algorithm to learn from the examples listed in \cref{sec:examples}. The success rate and model coefficient error statistics shown in \cref{fig:VanDerPol,fig:Lorenz,fig:Rossler} and \cref{fig:VanDerPol_ParamError,fig:Lorenz_ParamError,fig:Rossler_ParamError} are computed by running each algorithm multiple times at each noise level and data length, each time with a different noise seed.
While the author attempts to run these tests as many times as possible, owing to the computational cost of the endeavour, each algorithms are run different number of times: E-SINDy - 64 times, DySMHO - 64 times, modified-SINDy - 20 times and ODR-BINDy - 64 times.

\begin{table}[h!]
\begin{tabular}{p{0.07 \columnwidth}p{0.18 \columnwidth}p{0.18 \columnwidth}p{0.29 \columnwidth}p{0.18 \columnwidth}}
\hline \\
Example & Ensemble-SINDy & modified-SINDy & DySHMO & ODR-BINDy \\
\hline \\
Van Der Pol & \(\lambda\): 0.2, \texttt{nEnsembles}: 500,
\texttt{ensembleT}: 0.6, Weak Formulation: 9 pts, shape func: $\phi=(t^2-1)^2$  & \texttt{q}: 1, \(\lambda\):
0.05, \(\omega\): 0.9, \texttt{Nloop}:8, \texttt{N\_train}:15000,
\texttt{N\_SINDy\_Iter}: 15, \texttt{NormalizeLib}: false &
\texttt{nfe}: 80, \texttt{ncp}: 15, \texttt{data\_step}: 100,
\texttt{thresholding\_frequency}: 10, \texttt{thresholding\_tolerance}:
1, \texttt{horizon\_length}: 2.0,
\texttt{pre\_processing\_2} \texttt{significance}: 0.7 & \(\sigma_x\):
measuremet noise level prescribed, \(\sigma_{\partial t}\): 1e-2,
\(\sigma_Y\): 10, Finite Difference order: 4th \\
Lorenz63 & \(\lambda\): 0.2, \texttt{nEnsembles}: 100,
\texttt{ensembleT}: 0.6, Weak Formulation: 7 pts, shape func: $\phi=(t^2-1)^2$ & \texttt{q}: 3, \(\lambda\):
(0.1 if noise\(<10\%\), 0.15 otherwise), \(\omega\): 0.9,
\texttt{Nloop}:8, \texttt{N\_train}:15000, \texttt{N\_SINDy\_Iter}: 15,
\texttt{NormalizeLib}: false & \texttt{nfe}: 80, \texttt{ncp}: 5,
\texttt{data\_step}: 100, \texttt{thresholding\_frequency}: 10,
\texttt{thresholding\_tolerance}: 1, \texttt{horizon\_length}: 2.0,
\texttt{pre\_processing\_2} \texttt{significance}: 0.7 & \(\sigma_x\):
measuremet noise level prescribed, \(\sigma_{\partial t}\): 1e-3,
\(\sigma_Y\): 100, Finite Difference order: 6th \\
R{\"o}ssler & \(\lambda\): 0.05, \texttt{nEnsembles}: 500,
\texttt{ensembleT}: 0.65, Weak Formulation: 7 pts, shape func: $\phi=(t^2-1)^2$ & \texttt{q}: 1, \(\lambda\):
0.05, \(\omega\): 0.9, \texttt{Nloop}:8, \texttt{N\_train}:15000,
\texttt{N\_SINDy\_Iter}: 15, \texttt{NormalizeLib}: false &
\texttt{nfe}: 80, \texttt{ncp}: 5, \texttt{data\_step}: 100,
\texttt{thresholding\_frequency}: 10, \texttt{thresholding\_tolerance}:
1, \texttt{horizon\_length}: 5.0,
\texttt{pre\_processing\_2} \texttt{significance}: 0.7  & \(\sigma_x\):
measurement noise level prescribed, \(\sigma_{\partial t}\): 5e-3,
\(\sigma_Y\): 50, Finite Difference order: 6th \\
\hline
\end{tabular}
\caption{Hyperparameter settings used for each algorithms in the examples listed in \cref{sec:examples}.}
\label{tbl:hyperparameters}
\end{table}

\end{appendices}
\newpage
\bibliography{Library}

\end{document}